\documentclass[aps,prd,amssymb,eqsecnum,showpacs]{revtex4}
\usepackage{graphicx}
\usepackage{psfrag}

\begin{document}

\title{Constraint-preserving Sommerfeld conditions for the harmonic
Einstein equations}

\author{M.~C. Babiuc${}^{1}$, H-O. Kreiss${}^{2,3}$,
	Jeffrey Winicour${}^{1,3}$
       }
\affiliation{
${}^{1}$ Department of Physics and Astronomy \\
         University of Pittsburgh, Pittsburgh, PA 15260, USA \\
${}^{2}$NADA, Royal Institute of Technology, 10044 Stockholm, Sweden\\
${}^{3}$ Max-Planck-Institut f\" ur
         Gravitationsphysik, Albert-Einstein-Institut, \\
	 14476 Golm, Germany
	 }

\begin{abstract}

The principle part of Einstein equations in the harmonic gauge consists of a
constrained system of 10 curved space wave equations for the components of the
space-time metric.  A new formulation of constraint-preserving boundary
conditions of the Sommerfeld type for such systems has recently been proposed.
We implement these boundary conditions in a nonlinear 3D evolution code and
test their accuracy. 

\end{abstract}


\maketitle

\section{Introduction}

We present here the implementation and test results of a new formulation of
constraint-preserving Sommerfeld type boundary conditions for Einstein's
equations. The well-posedness of the initial-boundary value problem (IBVP)
for symmetric hyperbolic systems with maximally dissipative boundary
conditions can be established by the energy method~\cite{brown-book}. An
alternative technique, based on the principle of frozen coefficients,
Laplace-Fourier decomposition and the theory of pseudo-differential
operators, can be used to establish well-posedness in a generalized sense
even if the boundary conditions are not maximally dissipative or the system
is not symmetric hyperbolic~\cite{kr1970,green-book}. This theory has
recently been applied to formulate a well-posed, constraint-preserving IBVP
for the harmonic Einstein equations with boundary conditions of the
Sommerfeld type~\cite{wellposedsom}. In this paper, we show how these new
boundary conditions can be implemented in a finite-difference harmonic code
in which the Einstein equations are reduced to second order wave equations.
The test results presented here show that this new approach has potential
value for the computation of gravitational waves in a highly dynamical and
nonlinear regime. Although our application here is limited to test problems,
we expect these techniques to further the recent progress in the simulation
of black holes by harmonic evolution~\cite{pret,pret1,pret2,linsch}.

The first well-posed formulation of the IBVP for Einstein's equations was
presented in the pioneering work of Friedrich and Nagy~\cite{Friedrich98},
based upon a quite different formulation of the Einstein equations. The
underlying pseudo-differential theory and how it leads to a
constraint-preserving IBVP for the harmonic Einstein equations which is
well-posed in a generalized sense is described in~\cite{wellposedsom}. In
that work, the details were presented for the linearized Einstein equations
but it was explained how the general pseudo-differential theory extends 
well-posedness to the full nonlinear case. Subject to a certain inequality
which is necessary to establish the required estimates, there is
considerable freedom in the detailed form of the Sommerfeld-type
boundary conditions. 

In Sec.~\ref{sec:cpsbc} we describe how this new boundary treatment can be
implemented in a fully nonlinear code for the simplest choice of the boundary
conditions considered in~\cite{wellposedsom}. The resulting scheme is
attractive for numerical use. In a previous study~\cite{harl},
constraint-preserving boundary conditions for the harmonic Einstein equations
were based upon a combination of Dirichlet and Neumann conditions, which are
only marginally dissipative. The description of a traveling wave requires the
proper inhomogeneous Dirichlet or Neumann boundary data for the wave to pass
through the boundary. However, in numerical simulations, such inhomogeneous
Dirichlet or Neumann data can only be prescribed for the signal and the
numerical error is reflected by the boundary and accumulates in the grid. Test
results show that this can lead to poor performance in the simulation of
highly  dynamical and nonlinear solutions of Einstein's
equations~\cite{harm,babev}. For such computational purposes, it is more
advantageous to use a Sommerfeld condition which is strictly dissipative and
allows numerical error to leave the grid~\cite{bab}.

The numerical implementation of the boundary conditions, described in
Sec.~\ref{sec:numerical}, is carried out using two distinct approaches. One
approach is based upon {\em summation by parts} (SBP), which incorporates
semi-discrete versions of the conservation laws obeyed by the principle part
of the system. The stability of the finite difference scheme then follows
from a discretized energy argument. The second is based upon the embedded
boundary method, which is more easily applied to the case of a curved
boundary. Theorems regarding the stability of the embedded boundary method
for the second order wave equation have been given for the case of Dirichlet
and Neumann boundary conditions~\cite{krneum,krdir}. We expect that these
theorems can be extended to the strictly dissipative Sommerfeld case. 

We compare these two numerical implementations using the standardized
Apples with Apples linearized wave, gauge wave~\cite{mex1} and shifted gauge
wave~\cite{babev} tests, modified to include a boundary as documented
in~\cite{awa}. These tests allow a direct comparison of the SBP and embedded
boundary methods in a context where the boundary is aligned with the grid.
The test results are given in Sec.~\ref{sec:tests}.  

\section{Constraint-preserving Sommerfeld boundary conditions}
\label{sec:cpsbc}

Our results apply to the generalized harmonic formalism
including harmonic gauge source terms~\cite{Friedrich} and constraint
adjustments, as
described in ~\cite{constrdamp,babev}. Generalized
harmonic coordinates $x^\alpha=(t,x^i)=(t,x,y,z)$ are independent
solutions of the curved space scalar wave equation,
\begin{equation}
    \Box x^\mu  = \frac{1}{\sqrt{-g}}\partial_\alpha
           (\sqrt{-g}g^{\alpha\beta}\partial_\beta x^\mu) =-\hat \Gamma^\mu,
\end{equation}
with gauge source terms $\hat \Gamma^\mu(x^\alpha,g_{\alpha\beta})$ depending
on the coordinates and the metric. In terms of the connection
$\Gamma^\mu_{\alpha\beta}$, these harmonic conditions take the form
\begin{equation}
   {\cal C}^\mu :=\Gamma^\mu -\hat \Gamma^\mu =0.
\label{eq:constraints}
\end{equation}
where
\begin{equation}
     \Gamma^\mu = g^{\alpha\beta}\Gamma^\mu_{\alpha\beta}= 
              -\frac{1}{\sqrt {-g}}\partial_\alpha\gamma^{\alpha\mu}
\end{equation}
and $\gamma^{\mu\nu}=\sqrt{-g}g^{\mu\nu}$. 
These ${\cal C}^\mu$ are the constraints of the harmonic formulation.

Constraint adjustments of the form 
\begin{equation}
         A^{\mu\nu}={\cal C}^\rho A^{\mu\nu}_\rho 
   (x^\alpha,g^{\alpha\beta},\partial_\gamma g^{\alpha\beta}). 
\label{eq:constadj}
\end{equation} 
can  be introduced to modify the reduced system
of harmonic equations.
These reduced equations then take the form
\begin{equation}
   \tilde E^{\mu\nu} := G^{\mu\nu} -\nabla^{(\mu} {\cal C}^{\nu)} 
                 +\frac{1}{2}g^{\mu\nu}\nabla_\alpha {\cal C}^\alpha
		 + A^{\mu\nu} 
		 \label{eq:etilde},
\end{equation}
which are equivalent to Einstein equations $G^{\mu\nu}=0$ when the constraints
${\cal C}^\mu$ are satisfied. When the constraint adjustments and gauge source
terms vanish, they reduce to the standard harmonic reduction of the Einstein
tensor (see e.g.~\cite{wald,Friedrich}),
\begin{equation}
     E^{\mu\nu}:= G^{\mu\nu} -\nabla^{(\mu}\Gamma^{\nu)} 
                 +\frac{1}{2}g^{\mu\nu}\nabla_\alpha \Gamma^\alpha=0 .
		 \label{eq:e}
\end{equation}
The systems (\ref{eq:etilde}) and (\ref{eq:e}) have the same principle
part and  they both constitute a constrained system of quasilinear wave
equations with a well-posed Cauchy problem. In terms of the metric,
these quasilinear wave equations (\ref{eq:etilde}) can be put in the form
\begin{equation}
    2\sqrt{-g}{\tilde E}^{\mu\nu}=g^{\alpha\beta}\partial_\alpha\partial_\beta
         \gamma^{\mu\nu} + \hat S^{\mu\nu}=0, 
\label{eq:efc} 
\end{equation}
where $\hat S^{\mu\nu}$ are terms which do not contribute to the principle part.

The solutions of the generalized harmonic evolution system (\ref{eq:etilde})
are solutions of the Einstein equations provided the constraints ${\cal C}^\mu$ 
are satisfied. The Bianchi identities, applied to (\ref{eq:etilde}), imply
that ${\cal C}^\mu$ obeys the homogeneous wave equation
\begin{equation}
     \nabla^\alpha \nabla_\alpha {\cal C}^\mu +R^\mu_\nu {\cal C}^\nu 
                   -2\nabla_\nu ({\cal C}^\rho A^{\mu\nu}_\rho)=0.
     \label{eq:ahatconstr}
\end{equation}
The well-posedness of the Cauchy problem for (\ref{eq:ahatconstr}) enforces 
the
unique solution ${\cal C}^\mu =0$ in the domain of dependence of the initial
Cauchy hypersurface ${\cal S}$ provided the Cauchy data $\gamma^{\mu\nu}|_{\cal
S}$ and $\partial_t \gamma^{\mu\nu}|_{\cal S}$ satisfy ${\cal C}^\mu|_{\cal
S}=\partial_t {\cal C}^\mu|_{\cal S}=0$ via (\ref{eq:etilde}). It is
straightforward to show that these initial conditions are satisfied if the 
data on ${\cal S}$ satisfy the Hamiltonian and momentum constraints
$G^t_\mu=0$ and the initial condition ${\cal C}^\mu=0$.

In order to extend constraint preservation to the IBVP with boundary ${\cal B}$
it is sufficient to prescribe boundary conditions for $\gamma^{\mu\nu}$ which
imply a maximally dissipative homogeneous boundary condition for
${\cal C}^\mu$. In~\cite{harl,harm}, this was achieved by a combination of
Dirichlet and Neumann  boundary conditions on the various components of
$\gamma^{\mu\nu}$, which then induce a combination of homogeneous Dirichlet and
Neumann conditions on the components of ${\cal C}^\mu$. However, test results
showed that  Dirichlet and Neumann boundary conditions, which are only
borderline dissipative, were considerably less accurate than a strictly
dissipative Sommerfeld condition\cite{babev}. These test results were based upon
exact solutions for which the correct Sommerfeld data were known. Here we
consider another approach in which constrained Sommerfeld data may be applied
consistently in the absence of an exact solution. This Sommerfeld data for the
components of $\gamma^{\mu\nu}$ lead to the homogeneous Dirichlet condition
\begin{equation}
        {\cal C}^\mu|_{\cal S}=0
\end{equation}
on the constraints, which is sufficient to guarantee a constraint-preserving
well-posed IBVP. It is possible that a homogeneous Sommerfeld boundary
condition on the constraints would lead to better constraint preservation in
numerical applications. This requires a second differential order boundary
condition on the metric variables for which existing theory gives little
guidance. See~\cite{rinne} for a fuller discussion and some promising results.

Well-posedness depends only on the principle part of the quasilinear system
(\ref{eq:efc}). The pseudo-differential theory implies the {\it principle of
frozen coefficients} by which well-posedness for the nonlinear problem can be
established by treating the metric $g^{\alpha\beta}$ governing the wave
operator in (\ref{eq:efc}) as a constant. Thus the metric can be transformed by
a linear transformation into Minkowski form and without loss of generality it
suffices to establish well-posedness for the system of flat space wave
equations 
\begin{equation}
    \eta^{\alpha\beta}\partial_\alpha\partial_\beta \gamma^{\mu\nu} =0
\label{eq:emfc} 
\end{equation}
subject to the constraints (\ref{eq:constraints}). In this
local Minkowski frame, where $\sqrt{-g}=1$, the constraints reduce to
\begin{equation}
   {\cal C}^\mu :=-\partial_\nu \gamma^{\mu\nu}-\hat \Gamma^\mu =0.
\label{eq:tconstraints}
\end{equation}

We choose our local frame so that the $+x$-direction is the outward normal to
${\cal B}$ and the $x^a=(t,y,z)$-directions are tangent to ${\cal B}$.
We write $x^A=(y,z)$. The Sommerfeld boundary condition on a scalar field
$\Phi$ then takes the form
\begin{equation}
   (\partial_t +\partial_x)\Phi|_{\cal B} = q(x^a),
\end{equation}
where $q(x^a)$ represents the prescribed Sommerfeld data. Because
the Sommerfeld condition is strictly dissipative it leads to a well-posed
IBVP in the case of the scalar wave equation.

Constraint-preserving boundary conditions for the system of  linearized
Einstein equations (\ref{eq:emfc})-(\ref{eq:tconstraints}) can be expressed as
a hierarchy of Sommerfeld boundary conditions. There are numerous options in
this approach~\cite{wellposedsom} and here we consider the mathematically
simplest scheme.

First we require the 6 Sommerfeld boundary conditions
\begin{eqnarray}
  &(\partial_t +\partial_x)& \gamma^{AB} =q^{AB}(x^a) \label{eq:sombc1} \\
  &(\partial_t +\partial_x)&( \gamma^{tA}-\gamma^{xA} )
          =q^{tA}(x^a)-q^{xA}(x^a) \label{eq:sombc2} \\
  &(\partial_t +\partial_x)& ( \gamma^{tt}-2\gamma^{tx}+\gamma^{xx} )
       =q^{tt}(x^a)-2q^{tx}(x^a)+q^{xx}(x^a), \label{eq:sombc3}
\end{eqnarray} 
where the $q$'s are freely prescribed Sommerfeld data.
Next, the constraints are used to supply 4 additional
boundary conditions in the hierarchical order
\begin{eqnarray}
   &{\cal C}^A|_{\cal B}& =-\partial_t \gamma^{At}-\partial_x \gamma^{Ax}
      -\partial_B \gamma^{AB}  -\hat \Gamma^A (x^a) =0 \label{eq:cbc1} \\
   &{\cal C}^t|_{\cal B}&-{\cal C}^x|_{\cal B} =
      -\partial_t (\gamma^{tt}-\gamma^{xt})
       -\partial_x (\gamma^{xt}-\gamma^{xx}) 
    -\partial_B (\gamma^{tB}-\gamma^{xB})
           -\hat \Gamma^t (x^a)+\hat \Gamma^x (x^a)  =0 \label{eq:cbc2} \\\
   &{\cal C}^t|_{\cal B}& =-\partial_t \gamma^{tt}-\partial_x \gamma^{tx}
      -\partial_B \gamma^{tB}  -\hat \Gamma^t (x^a) =0. \label{eq:cbc3}
\end{eqnarray} 
By using (\ref{eq:sombc1})-(\ref{eq:sombc3}), the boundary conditions
(\ref{eq:cbc1})-(\ref{eq:cbc3}) can be re-expressed in the Sommerfeld form
\begin{eqnarray}
   {\cal C}^A|_{\cal B}& = &
      -\frac{1}{2}(\partial_t +\partial_x)( \gamma^{At}+ \gamma^{Ax})
       -\partial_t ( \gamma^{At}- \gamma^{Ax})
       -\partial_B \gamma^{AB} +\frac {1}{2}(q^{tA}-q^{xA}) -\hat \Gamma^A
       =0 
          \label{eq:scbc1} \\                      \nonumber \\
   {\cal C}^t|_{\cal B}-{\cal C}^x|_{\cal B} &=&
      -\frac{1}{2}(\partial_t +\partial_x)(\gamma^{tt}- \gamma^{xx})
     -\partial_t (\gamma^{tt}-2\gamma^{xt}+\gamma^{xx})     \nonumber \\
    &-&\partial_B (\gamma^{tB}-\gamma^{xB}) +\frac{1}{2}(q^{tt}-2q^{tx}+q^{xx})
           -\hat \Gamma^t+\hat \Gamma^x =0 
	     \label{eq:scbc2}  \\
   {\cal C}^t|_{\cal B}& =&
         -\frac{1}{2}(\partial_t +\partial_x)(\gamma^{tt}+ \gamma^{xx})
   -\partial_t (\gamma^{tt}-\gamma^{tx})+\frac{1}{2}(q^{tt}-2q^{tx}+q^{xx})
      -\partial_B \gamma^{tB}  -\hat \Gamma^t =0 .
       \label{eq:scbc3}
\end{eqnarray}
Equations (\ref{eq:sombc1})-(\ref{eq:sombc3}) and
(\ref{eq:scbc1})-(\ref{eq:scbc3}) form a hierarchical sequence of
inhomogeneous Sommerfeld boundary conditions in which the source terms
for (\ref{eq:scbc1})-(\ref{eq:scbc3}) are
provided by previous members in the hierarchy. In the linearized
problem, the boundary conditions  
(\ref{eq:sombc1})-(\ref{eq:sombc3}), along with compatible initial data that
satisfy the constraints, determine unique solutions of the wave
equations for $\gamma^{AB}$, $\gamma^{At}-\gamma^{Ax}$ and
$\gamma^{tt}-2\gamma^{tx}+\gamma^{xx}$. These then provide the source terms in 
(\ref{eq:scbc1}) and (\ref{eq:scbc2}), which determine unique solutions for 
$\gamma^{At}+\gamma^{Ax}$ and $\gamma^{tt}-\gamma^{xx}$. Finally, the source
terms in (\ref{eq:scbc2}) can be determined to provide a unique solution
for $\gamma^{tt}+ \gamma^{xx}$. 

The ten linearly independent Sommerfeld-type boundary conditions
(\ref{eq:sombc1})-(\ref{eq:sombc3}) and (\ref{eq:scbc1})-(\ref{eq:scbc3})
for the components of $\gamma^{\mu\nu}$ give rise to a unique solution of
the constrained linearized problem. However, it is important to emphasize
the following points:

\begin{enumerate}

\item  Proof of the well-posedness of the IBVP requires estimates on the
derivatives of the solution at the boundary. For that purpose, it is
required to apply the pseudo-differential theory to construct a
symmetrizer in Fourier-Laplace space, as described in~\cite{wellposedsom}.

\item The pseudo-differential theory allows well-posedness to be extended
locally in time to the nonlinear IBVP for the harmonic Einstein's
equations, where the relation $\gamma^{\mu\nu}=\sqrt{-g}g^{\mu\nu}$
converts (\ref{eq:efc}) into  a quasi-linear equation.  

\item Sommerfeld boundary conditions are strictly dissipative but they
completely remove all reflections only in highly idealized cases.

\item  The hierarchical structure of the boundary conditions, together with
their dissipative property, is a very good heuristic procedure to formulate
a stable finite difference approximation but by itself does not provide a
proof of stability.

\item In the electromagnetic case, our approach gives rise to
constraint-preserving Sommerfeld-type boundary conditions on the vector
potential which can be given a physical interpretation in terms of the
Poynting vector~\cite{wellposedsom}. An analogous interpretation does not
exist in the gravitational case. When $q^{AB}$ has vanishing trace-free 
part, the boundary condition (\ref{eq:sombc1}) implies that the outgoing
null hypersurfaces emanating from the boundary have vanishing shear. This
is related to the vanishing of incoming radiation but in a very
gauge dependent way. Friedrich and Nagy~\cite{Friedrich98} stress this same
caveat concerning their boundary conditions, which specify the
Newman-Penrose Weyl component $\Psi_0$ associated with the outgoing null
hypersurfaces.

\item In~\cite{wellposedsom}, well-posedness of the IBVP was established
for the special case where the boundary $x=0$ contains the timelike
normal to the initial Cauchy hypersurface $t=0$. More generally, the
initial Cauchy hypersurface is given by $T=t-k x$ so that the initial data
is different from the initial data at $t=0$. Well-posedness of this more
general IBVP can be reduced to well-posedness of the special case by first
considering a pure Cauchy evolution from $T=0$ to $t=0$. The general case
corresponds to a boundary that is a ``moving'' with respect to the Cauchy
hypersurface. This introduces a shift term in the wave operator.  In the
next section, we give the details of how to translate the conditions
(\ref{eq:sombc1})-(\ref{eq:sombc3}) and (\ref{eq:scbc1})-(\ref{eq:scbc3})
into a nonlinear computational scheme for a general metric which is not in
the shift-free Minkowski form. 

\end{enumerate}

\section{Numerical implementation}
\label{sec:numerical}

Our implementation is based upon the Abigel code~\cite{harl,harm},
in which the harmonic system
(\ref{eq:efc} ) is integrated in the first order in time form
\begin{eqnarray}
    \partial_t \gamma^{\mu\nu} &=& T^{\mu\nu} 
\label{eq:evgamma} \\
    \partial_t T^{\mu\nu} &=& 
       F^{\mu\nu}(\gamma,\partial_i \gamma, T, \partial_i T).
\label{eq:evT}
\end{eqnarray}
Here $x^i=(x,y,z)$ and $\partial_i$ denotes spatial derivatives. The code is
based upon an explicit, second order accurate finite-difference scheme.
Introduction of a spatial grid and finite difference approximations for the
spatial derivatives reduces (\ref{eq:evgamma})-(\ref{eq:evT}) to a large set of
ordinary differential equations (method of lines), which are evolved with a
fourth order Runge-Kutta integration. Details of the finite difference
approximations are given in~\cite{harl,harm}, where the system
(\ref{eq:evgamma})-(\ref{eq:evT}) was expressed in flux conservative form and
summation by parts (SBP) was used to apply the boundary conditions, which
enforced the semi-discrete version of the conservation laws obeyed by the
principle part The tests in this paper have been carried out with the $\hat W$
form of the algorithm described in~\cite{harm}, in which certain nonlinear
coefficients are approximated by their averages between grid points. We also
add artificial dissipation to (\ref{eq:evgamma}) and (\ref{eq:evT}) by the
modifications
\begin{eqnarray}
    \partial_t \gamma^{\mu\nu} &\rightarrow& \partial_t \gamma^{\mu\nu}
      + \epsilon_\gamma {\cal D}^4 \gamma^{\mu\nu}
\label{eq:dissgamma} \\
    \partial_t T^{\mu\nu} &\rightarrow& \partial_t T^{\mu\nu}
        + \epsilon_T {\cal D}^4 T^{\mu\nu} ,
\label{eq:dissT}
\end{eqnarray}
where ${\cal D}^2$ is the SBP approximation for the
Laplace operator.

The SBP approach requires that the boundary be aligned with the
numerical grid. Here we also consider implementation of the boundary conditions
by means of the embedded boundary method~\cite{krneum,krdir}, which is
applicable even when the boundary is not aligned with the grid.

\subsection{The Sommerfeld boundary data}
\label{sec:somm}

We describe the implementation of constraint-preserving Sommerfeld
boundary conditions for the fully nonlinear harmonic system (\ref{eq:etilde}).
For purpose of discussion, we locate the boundary ${\cal B}$ at $x=const$ with
the outer normal in the $+x$-direction. We denote the coordinates intrinsic
to the boundary by $x^a=(t,y,z) = (t,x^A)$.  We introduce an orthonormal tetrad
$(T^\mu, X^\mu, Y^\mu, Z^\mu)$, oriented at ${\cal B}$  so that
$X_\mu=(0,1/\sqrt{g^{xx}},0,0)$ is the unit outward normal and
$T^\mu=(1/\sqrt{-g_{tt}},0,0,0)$ is a unit vector in the evolution direction.
In addition to the standard requirements of the IBVP that the Cauchy
hypersurfaces be spacelike and that the boundary be timelike, we require that
$T^\mu$ be timelike, i.e. that the evolution be subluminal. 

The Sommerfeld data for $\gamma^{\mu\nu}$ consist of
\begin{equation}
    K^\alpha\partial_\alpha \gamma^{\mu\nu} =q^{\mu\nu},
\end{equation}
where $K^\mu=T^\mu+X^\mu$ and $K^\mu K_\mu =0$. The description
of the free and constrained components of $q^{\mu\nu}$ is algebraically more
complicated than for the Minkowski case in Sec.~\ref{sec:cpsbc}. For this
reason, it is useful to introduce the projection operators
\begin{equation}
   M^\nu_\mu=\delta^\nu_\mu +T_\mu T^\nu -X_\mu X^\nu=Y_\mu Y^\nu+Z_\mu Z^\nu,
\label{eq:mproj}
\end{equation}
which projects vectors into the $(Y^\mu,Z^\mu)$ spatial-plane tangent to the
boundary, and
\begin{equation}
     P^\nu_\mu=\delta^\nu_\mu +\frac{1}{2} L_\mu K^\nu,
\label{eq:proj}
\end{equation}
where $L^\mu=T^\mu -X^\mu$ with $K^\mu L_\mu =-2$, which
projects vectors into the null 3-space spanned by $(L^\mu,Y^\mu,Z^\mu)$. 
We have $P^\nu_\mu K^\mu=0$ and $P^\nu_\mu L_\nu =0$, i.e. $P^\nu_\mu$
projects forms $W_\mu$ into the subspace orthogonal to $K_\mu$. We have
\begin{equation}
   P^\nu_\mu= M^\nu_\mu -\frac{1}{2} K_\mu L^\nu.
\end{equation}
The freely prescribed
Sommerfeld data corresponding to (\ref{eq:sombc1})-(\ref{eq:sombc3})
consist of the projection
\begin{equation}
    Q^{\mu\nu}= P^\mu_\alpha P^\nu_\beta q^{\alpha\beta}.
\end{equation}
The constrained data consist of
$Q^\mu=P^\mu_\alpha L_\beta q^{\alpha\beta}$, analogous to
(\ref{eq:scbc1})-(\ref{eq:scbc2}); and $Q=L_\alpha L_\beta
q^{\alpha\beta}$, analogous to (\ref{eq:scbc3}). In order to determine the
constrained data, we use the identity
\begin{equation}
   \partial_\alpha \gamma^{\alpha\nu}= -\frac{1}{2} L_\alpha q^{\alpha\nu} 
             + P^\rho_\alpha \partial_\rho  \gamma^{\alpha\nu}.    
\end{equation}
The constraints then imply
\begin{equation}
  -\sqrt{-g} \hat \Gamma^\nu= -\frac{1}{2} L_\alpha q^{\alpha\nu} 
             + P^\rho_\alpha \partial_\rho  \gamma^{\alpha\nu}
\end{equation}
so that
\begin{equation}
    Q^\mu =   2 P^\mu_\alpha  P^\rho_\beta \partial_\rho \gamma^{\alpha\beta}
     +2\sqrt{-g}P^\mu_\alpha \hat \Gamma^\alpha
\label{eq:qa}
\end{equation}
and 
\begin{equation}
   Q =  2 L_\alpha  P^\rho_\beta \partial_\rho \gamma^{\alpha\beta}
     +2\sqrt{-g}  L_\alpha \hat \Gamma^\alpha.
\label{eq:q}
\end{equation}
The full set of Sommerfeld data consist of 
\begin{equation}
  q^{\mu\nu} = Q^{\mu\nu} - Q^{(\mu}K^{\nu )} + \frac{1}{4} Q K^\mu K^\nu.
\label{eq:qQ}
\end{equation}

As in (\ref{eq:scbc1})-(\ref{eq:scbc3}), the derivatives normal to the boundary
occurring in (\ref{eq:qa}) and (\ref{eq:q}) can be eliminated via the
identities 
\begin{equation}
  P^\mu_\alpha  P^\rho_\beta \partial_\rho \gamma^{\alpha\beta} =
      X_\alpha Q^{\mu\alpha}
         -P^\mu_\alpha K_\beta T^\rho \partial_\rho\gamma^{\alpha\beta}
     +P^\mu_\alpha M^\rho_\beta \partial_\rho\gamma^{\alpha\beta}
\label{eq:qaex}
\end{equation}
and
\begin{equation}
     L_\alpha  P^\rho_\beta \partial_\rho \gamma^{\alpha\beta}=
      \frac{1}{2} K_\alpha Q^\alpha
         - L_\alpha K_\beta T^\rho \partial_\rho\gamma^{\alpha\beta}
      + L_\alpha M^\rho_\beta \partial_\rho\gamma^{\alpha\beta}.
\label{eq:qex}
\end{equation}
Combined with (\ref{eq:qa}) and (\ref{eq:q}), these equations determine the
constrained boundary data $Q^\mu$ and $Q$ in terms of quantities intrinsic to
the boundary,
\begin{equation}
    Q^\mu =   2 \bigg (  \frac{1}{2}K_\alpha Q^{\mu\alpha}
         -P^\mu_\alpha K_\beta T^\rho \partial_\rho\gamma^{\alpha\beta}
     +P^\mu_\alpha M^\rho_\beta \partial_\rho\gamma^{\alpha\beta}
     +\sqrt{-g}P^\mu_\alpha \hat \Gamma^\alpha  \bigg)
\label{eq:Qa}
\end{equation}
\begin{equation}
   Q =  2\bigg ( \frac{1}{2} K_\alpha Q^\alpha
         - L_\alpha K_\beta T^\rho \partial_\rho\gamma^{\alpha\beta}
      + L_\alpha M^\rho_\beta \partial_\rho\gamma^{\alpha\beta}
       +\sqrt{-g}  L_\alpha \hat \Gamma^\alpha \bigg ).
\label{eq:Q}
\end{equation}

\subsection{The boundary update algorithm}

In Sec.~\ref{sec:somm}, we have described how the boundary values of
$\gamma^{\mu\nu}$, $T^{\mu\nu}=\partial_t \gamma^{\mu\nu}$ and the free
Sommerfeld data $Q^{\mu\nu}$ at time $t$ determine the full Sommerfeld data
$q^{\mu\nu}$ at time $t$. We now consider how these quantities are updated.
The update of the boundary values of $\gamma^{\mu\nu}$
to the next time step $t+\Delta t$ is straightforward via (\ref{eq:evgamma})).
The update of $T^{\mu\nu}$ is more complicated. Here we use two distinct
algorithms, the SBP algorithm and the embedded boundary algorithm, for updating
the boundary values of $T^{\mu\nu}$. This provides two competitive update
algorithms, which are compared in test problems in Sec.~\ref{sec:tests}.

{\bf SBP algorithm}. This is the algorithm described in~\cite{harm}. The
evolution equation (\ref{eq:evT}) is applied at the boundary points to update
$T^{\mu\nu}$, with field values at the resulting ghost points eliminated by the
boundary condition in a manner that enforces discrete conservation laws. Full
details for the cases of Dirichlet, Neumann and Sommerfeld boundary conditions
are given in~\cite{harm}. The only new ingredient here is the use of
constraint-preserving Sommerfeld conditions. However, because the constrained
Sommerfeld data $Q^\mu$ and $Q$ can be updated {\em after} the update of
$\gamma^{\mu\nu}$ and $T^{\mu\nu}$, there is no essential change in the
numerical algorithm.	

{\bf Embedded boundary algorithm}.

The SBP algorithm uses the fact that the boundary is aligned with the grid, so
that boundary points are grid points. In the case of a spherical boundary and a
Cartesian grid, this is generally not the case unless multi-block techniques
are used.  An alternative approach for treating a curved boundary with a
Cartesian grid is the {\em embedded boundary} method, in which field values at
ghost points are updated from interpolations using the boundary data, as
opposed to applying the evolution equation. 

Assume that we are given the values of
$\gamma^{\mu\nu}$ and $T^{\mu\nu}$ at time $t$ and the free data $Q^{\mu\nu}$
at all times. As above, this determines the full constrained Sommerfeld data
$q^{\mu\nu}$ at time $t$. After updating the boundary values of
$\gamma^{\mu\nu}$ to time $t+\Delta t$, we update the boundary values
$T^{\mu\nu}$ using
\begin{equation}
     q^{\mu\nu} = K^\alpha \partial_\alpha \gamma^{\mu\nu} =
               K^t T^{\mu\nu} + K^i \partial_i \gamma^{\mu\nu},
\label{eq:Tupdate}
\end{equation}
where the spatial derivatives $K^i\partial_i \gamma^{\mu\nu}$ at the boundary 
are determined to by an interpolation scheme (see below). When all components
of the Sommerfeld data $q^{\mu\nu}$ are supplied analytically, this provides
updated  boundary values for $T^{\mu\nu}$, which we refer to as the {\it
analytic} version of the embedded boundary algorithm.

In the absence of an analytic solution, only the unconstrained components of
the Sommerfeld data $Q^{\mu\nu}$ can be prescribed freely. The update of
$T^{\mu\nu}$ is then first carried out for those components determined by
the free data, i.e.
\begin{equation}
          \tau^{\mu\nu}=P^\mu_\alpha P^\nu_\beta T^{\alpha\beta},
\end{equation}
for which (\ref{eq:Tupdate}) gives
\begin{equation}
     Q^{\mu\nu} = K^t \tau^{\mu\nu}
              + P^\mu_\alpha P^\nu_\beta K^i \partial_i \gamma^{\alpha\beta}.
\label{eq:taumnupdate}
\end{equation}

In order to update the boundary values of the remaining components
of $T^{\mu\nu}$, we must first update the constrained Sommerfeld data
$Q^\mu$ and $Q$ by expressing all time derivatives on the right hand sides
of (\ref{eq:Qa}) and (\ref{eq:Q}) in terms of previously updated
quantities. 
From (\ref{eq:Qa}), we obtain 
\begin{equation}
    Q^\mu =   2 \bigg ( \frac{1}{2}K_\alpha Q^{\mu\alpha}
         -\frac {1}{\sqrt{-g_{tt}}} K_\alpha \tau^{\mu\alpha}
     +M_\alpha^t \tau^{\mu\alpha}
    +P^\mu_\alpha M_\beta^i \partial_i \gamma^{\alpha\beta}
     +\sqrt{-g}P^\mu_\alpha \hat \Gamma^\alpha  \bigg) .
\label{eq:Qas}
\end{equation}
This determines 
\begin{equation}
          \tau^\mu=P^\mu_\alpha L_\beta T^{\alpha\beta} 
\end{equation}
through
\begin{equation}
     Q^\mu = K^t \tau^\mu
              + P^\mu_\alpha L_\beta K^i \partial_i \gamma^{\alpha\beta}.
\label{eq:taumupdate}
\end{equation}

Similarly, from (\ref{eq:Q}), we next obtain
\begin{equation}
   Q =  2\bigg ( \frac{1}{2} K_\alpha Q^\alpha 
      -\frac {1}{\sqrt{-g_{tt}}} K_\alpha \tau^\alpha
       +M_\alpha^t \tau^\alpha
       + L_\alpha M_\beta^i \partial_i\gamma^{\alpha\beta}
       +\sqrt{-g}  L_\alpha \hat \Gamma^\alpha \bigg ).
\label{eq:Qs}
\end{equation}
This determines 
\begin{equation}
          \tau=L_\alpha L_\beta T^{\alpha\beta} 
\end{equation}
through
\begin{equation}
     Q = K^t \tau
              + L_\alpha L_\beta K^i \partial_i \gamma^{\alpha\beta}.
\label{eq:tauupdate}
\end{equation}
These pieces allow us to construct
\begin{equation}
    T^{\mu\nu} = \tau^{\mu\nu} - \tau^{(\mu}K^{\nu )} 
         + \frac{1}{4} \tau K^\mu K^\nu.
\label{eq:Tconstruct}
\end{equation}

In carrying out this update of $T^{\mu\nu}$ it is
necessary to compute the boundary values of the spatial derivatives $K^i
\partial_i \gamma^{\mu\nu}$ which appear in (\ref{eq:Tupdate}).  This is
accomplished by an interpolation scheme patterned after the embedded boundary
treatment for a Neumann condition given in~\cite{krneum}. In this scheme, the
value of $K^i\partial_i \gamma^{\mu\nu}$ at a boundary point $B$, with grid
values $x^{i}_B=(I_B,J_B,K_B)h$, is obtained from the 1-dimensional
Lagrange polynomial for $\gamma^{\mu\nu}$ determined by B and $M$ interior
points, $C_m$ $(1\le m \le M)$, lying along the curve
\begin{equation}
      x^i =x^{i}_B + \frac {K^i} {K^x} (x-x_B) 
\label{eq:curve}
\end{equation}
normal to the boundary. We choose $(x_B-x_{C_m})=mh$ so that the points $C_m$
lie in $(y,z)$ planes through the $x$-grid. The values of $\gamma^{\mu\nu}$ at
the points $C_m$ are obtained by 2D interpolations in those planes. We use a 2D
Lagrange interpolant based upon a stencil of grid points which is symmetrical
about the point  $(y_B,z_B)$. In order to avoid extrapolation, the size of the
stencil has to be adjusted to the size of ${K^i}/{K^x}$. For all 2D test cases
considered in Sec.\ref{sec:tests}, it suffices to use a $3\times 3$ (or larger)
stencil  in the planes determined by $C_m$.

Note that this scheme differs from the straightforward computation of $K^i
\partial_i \gamma^{\mu\nu}$ by centered differencing in the $y$ and $z$
directions and one-sided differencing in the $x$ direction, which would give a
qualitatively different and less accurate approximation. However, in the simple
case where $K^y=K^z=0$, this interpolation scheme does reduce to approximating
$K^i \partial_i \gamma^{\mu\nu}=K^x\partial_x \gamma^{\mu\nu}$ by a 
one-sided finite difference at $B$. For a second order accurate scheme, only 3
points $C_m$ $(M=3)$ are necessary to construct the Lagrange
polynomial approximating (\ref{eq:curve}). However, with a second order
accurate interior evolution algorithm, the error would then be largest at the
boundary because of the use of a 1-sided derivative. On the other hand, it
is simple to use more points if higher accuracy at the boundary is desired. In
this respect, the embedded boundary algorithm is more flexible than the SBP
algorithm, for which the order of accuracy of the boundary algorithm is coupled
with the order of accuracy of the interior algorithm. In the embedded case, for
any internal accuracy, the order of accuracy of the boundary condition can be
made high as desired at little additional computational expense. In the tests
results shown in the next section, we use $M=5$. We also add a corrector
step to the boundary update algorithm to improve accuracy.

\section{Tests}
\label{sec:tests}

We conduct three boundary tests which have been proposed for the
AppleswithApples (AwA) test suite~\cite{awa}. These tests extend the original
linearized wave, gauge wave and shifted gauge wave tests with periodic
boundaries, i.e. a 3-torus $T^3$ without boundary, to include non-trivial
boundaries by opening up the $x$-axis of the 3-torus to form a manifold with
smooth $T^2$ boundaries at $x=\pm.5$. In this way, the tests avoid the
complication of sharp boundary points. The corresponding metrics are
\begin{itemize}
\item Linearized wave:
\begin{equation}
  ds^2=-dt^2 +dx^2+(1+H)dy^2+(1-H)dz^2,
  \label{eq:pw}
\end{equation}

\item Gauge wave:
\begin{equation}
  ds^2=(1-H)(-dt^2 +dx^2)+dy^2+dz^2,
  \label{eq:gw}
\end{equation}

\item Shifted gauge wave: 
\begin{equation}
  ds^2=- dt^2 +dx^2+dy^2+dz^2 +H k_\alpha k_\beta dx^\alpha dx^\beta,
  \label{eq:sgw}
\end{equation}

\end{itemize}

where in all cases
\begin{equation}
  \label{eq:flatgaugewaveHfn}
  H = H(x-t)= A \sin \left( \frac{2 \pi (x - t)}{d} \right),
\end{equation}
and
\begin{equation}
      k_\alpha=\partial_\alpha (x-t)=(-1,1,0,0).
\end{equation}
These metrics describes sinusoidal traveling waves of amplitude $A$ propagating
along the $x$-axis. Two dimensional features are tested by 
rotating the coordinates according to 
\begin{equation}
  \label{eq:flatgaugewave1to2d}
   x = \frac{1}{\sqrt{2}}(x^\prime - y^\prime), \qquad
    y = \frac{1}{\sqrt{2}}(x^\prime + y^\prime) \, .
\end{equation}
which produces a wave propagating along the diagonal. 

The linearized wave test is run with an amplitude  $A=10^{-8}$. It is most
efficient for revealing problems arising from nonlinearity to run the gauge wave
and shifted gauge wave tests with amplitude $A=0.5$. In some cases we also run
with smaller amplitudes in the range $A=.01$ to $A=.1$ (the original AwA
specifications) to reveal the emergence of nonlinear features. In all other
respects, we retain the original AwA specifications:  
\begin{itemize}
 \item Wavelength: $d=1$ in the 1D simulation and
          $d'=1/\sqrt{2}$ in the 2D simulation.
\item Simulation domain:
\begin{center}
\begin{tabular}{rllll}
  1D:& $\quad  x \in [-0.5, +0.5],$ & $\quad y = 0,$ &$ \quad z = 0,$
  & $ \quad d=1$ \\
  diagonal: & $\quad x \in [-0.5, +0.5], $ & 
    $\quad y \in [-0.5, +0.5], $ & $\quad z=0,$ & $\quad d'=\sqrt{2}$
\end{tabular}
\end{center}
\item Grid: $x_n = -0.5 + n dx, \quad n=0,1\ldots 50\rho,
  \quad dx=dy=dz=1/(50\rho), \quad \rho = 1, 2, 4 $
\item Time step: $dt = dx/4 = 0.005 / \rho$ .
\end{itemize}
The grids have $N=50 \rho =(50,100,200)$ zones. (At least 50 zones are required
to lead to reasonable simulations for more than 10 crossing times.)  The 1D
tests are carried out for $t=1000$ crossing times, i.e.~$2\times10^5\rho$
time steps, and the 2D tests for 100 crossing times.

As an example of how the Sommerfeld boundary data is prescribed, consider the
1D shifted gauge wave. The Sommerfeld operator at the right boundary $x=+.5$ is
given by $K^\alpha \partial_\alpha =\sqrt{1-H}(\partial_t+\partial_x)$ and the
corresponding Sommerfeld data vanishes ($q^{\mu\nu}=0$).  At the left boundary
$x=-.5$, where the outward normal is in the $(-x)$-direction, the Sommerfeld
operator is
\begin{equation}
    K^\alpha \partial_\alpha = (T^\alpha-X^\alpha)\partial_\alpha
            =\frac {1+H}{\sqrt{1-H}}\partial_t- \sqrt{1-H}\partial_x
\end{equation}
and the
resulting non-vanishing components of Sommerfeld data are
\begin{equation}
    q^{tt}=q^{tx}=q^{xx}=-\frac{2}{\sqrt{1-H}}\partial_t H(x-t).
\label{eq:sgwdata}
\end{equation}
With this inhomogeneous Sommerfeld data, the wave enters through the boundary
at $x=-.5$, propagates across the grid and exits through the boundary at
$x=+.5$.

At the left boundary $x=-.5$, the inward null direction $L_\mu = T_\mu+X_\mu$,
which enters the projection operator (\ref{eq:proj}), satisfies $L_\mu
q^{\mu\nu}=0$. Consequently, the non-vanishing components of the free data
$Q^{\mu\nu}$ are also given by (\ref{eq:sgwdata}) and the constrained components
vanish, i.e. $Q^\mu=Q=0$. Of course, in a simulation using the constraint
preserving Sommerfeld algorithm, numerical error gives rise to  non-vanishing
values for $Q^\mu$ and $Q$. Thus it makes a difference at the numerical level
whether $Q^\mu$ and $Q$ are given their analytic values (zero) or their values
produced by the constraint-preserving algorithm. These alternatives will be
compared in carrying out the tests to provide a measure of the efficacy of the
constraint-preserving algorithm.

Thus for each of the two boundary algorithms (SBP and embedded boundary) we run
the tests (i) with all 10 components of Sommerfeld data $q^{\mu\nu}$ provided by
the analytic solution and (ii) with the 6 free components of Sommerfeld data
$Q^{\mu\nu}$ provided by the analytic solution and the remaining components
$Q^\mu$ and $Q$ provided by the constraint-preserving algorithm. We distinguish
the corresponding tests results by the labels ASBP and AEMB, respectively, for
the SBP and embedded algorithms with fully analytic data; and CSBP and CEMB for
the corresponding tests with constraint-preserving data.

In the 1D gauge wave tests, the curves (\ref{eq:curve}) normal to the boundary
pass through the grid points on the $x$-axis, so that the interpolations
required for the embedded boundary algorithm are trivial. The 2D gauge wave and
shifted gauge wave both provide a non-trivial test of the interpolation scheme.
For the case of the shifted gauge wave, the curves (\ref{eq:curve}) at the
boundaries $x_B=\pm.5$ are given by
\begin{equation}
        y =y_B + \frac {K^y} {K^x} (x-x_B) .
\label{eq:sgwcurve}
\end{equation}   

Harmonic gauge forcing terms were not found to be effective in the
boundary-free gauge wave tests and we have not included them in the present
tests. (Gauge forcing is important in spacetimes where harmonic coordinates
become pathological, e.g. the standard $t$-coordinate in Schwarzschild
spacetime is harmonic but singular at the horizon.) Constraint adjustments were
not necessary except in trying to stabilize the constrained SBP shifted gauge
wave runs. 

We use the $\ell_\infty$ norm to measure the error
\begin{equation}
       {\cal E}(\Phi) =||\Phi_\rho-\Phi_{ana}||_\infty
\end{equation}
in a grid function $\Phi_\rho$ with known analytic value $\Phi_{ana}$. We
measure the convergence rate at time $t$
\begin{equation}
    r(t) = \log_2 \big (
   \frac{||\Phi_2-\Phi_{ana}||_\infty}{||\Phi_4-\Phi_{ana}||_\infty} \big ),
\end{equation}
using the $\rho=2$ and $\rho=4$ grids ($N=100$ and $N=200)$. (The $\rho=1$
grid is not very accurate and is mainly used for debugging.) For convergence
studies it is also useful to graph the rescaled error
\begin{equation}
      {\cal E_\rho} =\frac{\rho^2}{16}||\Phi_\rho-\Phi_{ana}||_\infty,
\end{equation}
which is normalized to the $\rho=4$ grid.

\subsection{Linearized wave tests}

\begin{figure}[hbtp]
  \centering
  \psfrag{xlabel}{x}
  \psfrag{ylabel}{$g_{yy}-1$}
  \includegraphics*[width=7cm]{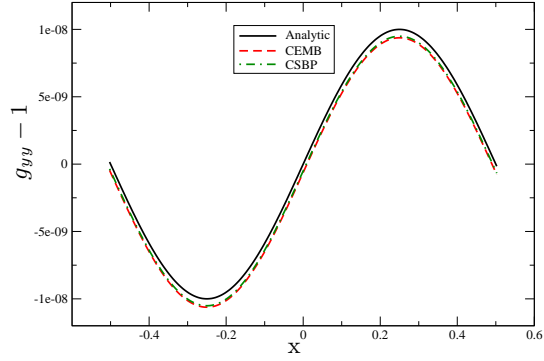}
  \caption{The graphs compare the performance of the CEMB and CSBP 
          algorithms for the 1D linearized wave test on the $\rho=4$ grid
	  with the analytic solution.
          The  CEMB and CSBP snapshots of $g_{yy}(x)-1$, shown at $t=1000$,
          are indistinguishable and show only a small amplitude discrepancy
	  with the analytic solution.
          }
  \label{fig:1Dlwlt}
\end{figure}

Table~\ref{tab:lwconv1D} shows the convergence rates of the error in $g_{yy}$
for the 1D linearized wave, measured at $t=10$ and $t=50$ crossing times. Clean
second order convergence is maintained for all four algorithms, irrespective of
whether the complete Sommerfeld data is supplied from the analytic solution
(ASBP and AEMB), or whether it is constrained (CSBP and CEMB). At 1000 crossing
times, the four algorithms continue to give excellent agreement with the
analytic solution. The graphs in Fig.~\ref{fig:1Dlwlt} show excellent phase
agreement and a small difference in amplitude at $t=1000$ in the comparison
between the analytic solution for $g_{yy}(x)$ and the results the CEMB and CSBP
results. On the scale of Fig.~\ref{fig:1Dlwlt}, the graphs for the AEMB and
ASBP algorithms are indistinguishable from the analytic solution.

\begin{table}[htbp]
  \begin{center}
    \begin{tabular}[c]{|c|c|c|}
\hline
  ALGORITHM &
  t=10 &
  t=50 
   
    \\ \hline \hline
AEMB		 & $2.04$ & $2.05$ 
\\ \hline
ASBP		 & $1.99$ & $1.93$ 
\\ \hline
CEMB		 & $2.04$ & $1.97$ 
\\ \hline
CSBP		 & $1.99$ & $2.03$ 
\\ \hline
\end{tabular}
    \caption{Convergence rates of ${\cal E}(g_{yy})$ for the
             1D linearized wave test, amplitude $A=10^{-8}$.}
    \label{tab:lwconv1D}
  \end{center}
\end{table}

\begin{figure}[hbtp]
  \centering
  \psfrag{xlabel}{x}
  \psfrag{ylabel}{$g_{yy}-1$}
  \includegraphics*[width=7cm]{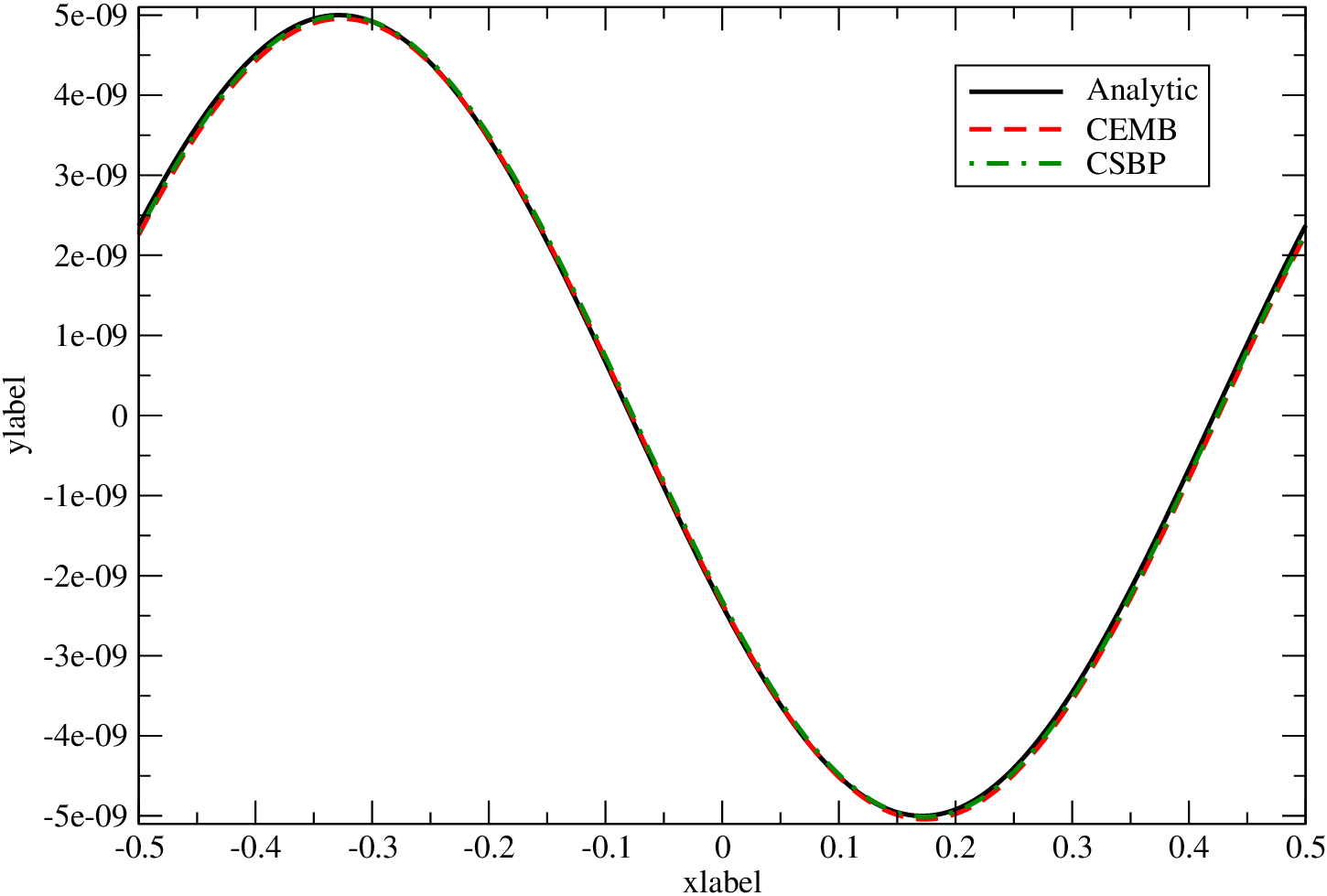}
  \includegraphics*[width=7cm]{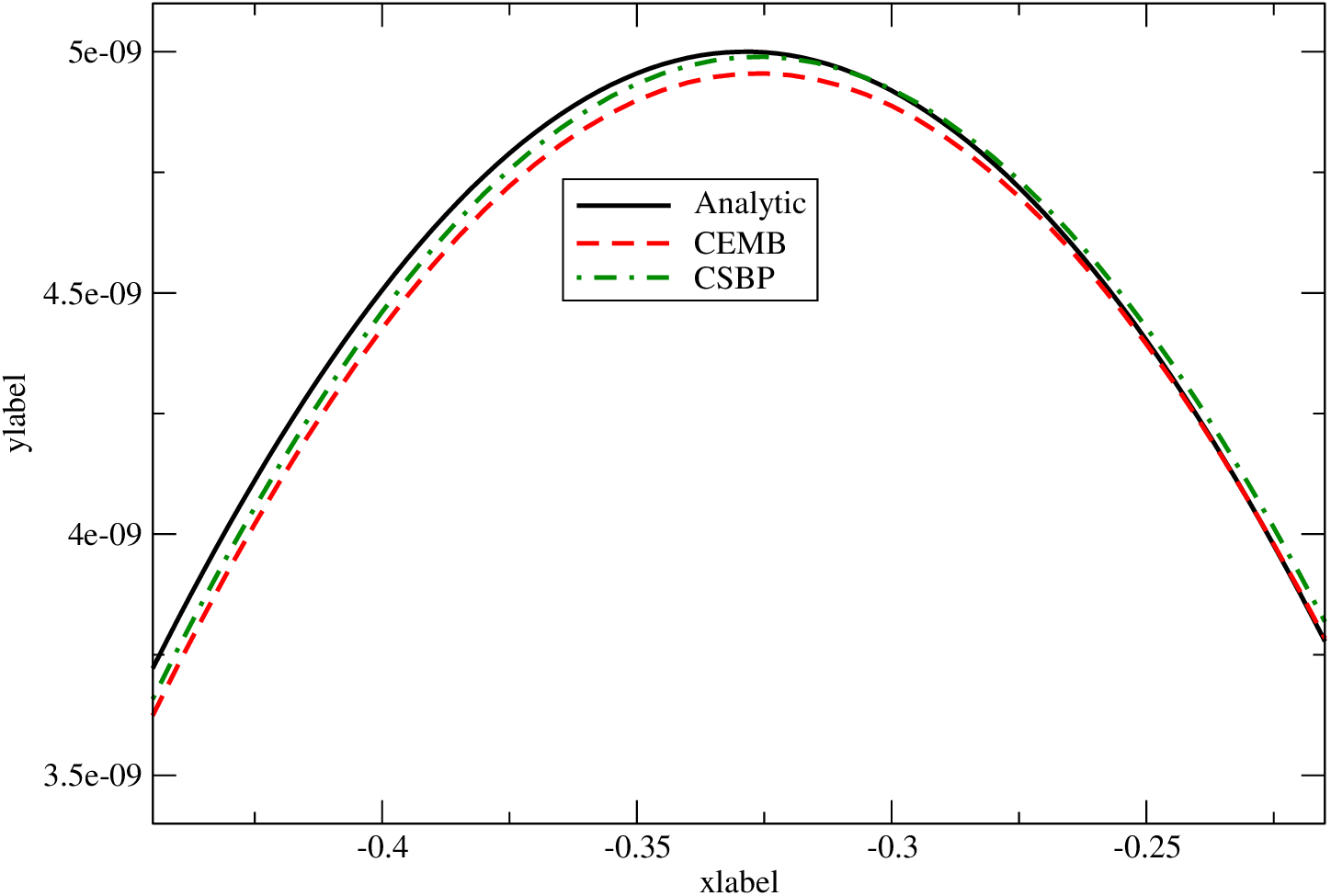}
  \caption{Snapshots of $g_{yy}(x)-1$
  for the 2D linear wave tests obtained at $t=100$, setting $y=0$. 
  On the left plot, the snapshots of $g_{yy}-1$ for the
  CEMB and CSBP algorithms are compared with the analytic solution. On this
  scale the errors are almost imperceptible and a zoom is given in the
  the right plot.}
  \label{fig:2Dlwlt}
\end{figure}

Table~\ref{tab:lwconv2D} shows the convergence rates of the error in $g_{yy}$
for the 2D linearized wave test, measured at $t=1$, $t=5$ and $t=10$. Second
order convergence is cleanly maintained for the AEMB and ASBP algorithms.
The convergence rates of the CEMB and CSBP show some
deterioration from second order at $t=10$ as the truncation error
from the boundary algorithms accumulates. Nevertheless, at $t=100$, the
2D results for all the algorithms remain in excellent
agreement with the analytic solution. The projection operators used in the
constrained CSBP and CEMB algorithms introduce the small errors shown in
Fig.~\ref{fig:2Dlwlt}, where snapshots of $g_{yy}-1$ are graphed for
$y=0$ and $t=100$, using the $\rho=4$ grid. Overall, the linearized wave
tests show that both the SBP and embedded algorithms give excellent results
for either fully analytic or constrained Sommerfeld boundary conditions. 

\begin{table}[htbp]
  \begin{center}
    \begin{tabular}[c]{|c|c|c|c|}
\hline
  ALGORITHM &
  t=1 &
  t=5 &
  t=10 
   
    \\ \hline \hline
AEMB		 & $1.98$ & $2.00$ & $1.97$ 
\\ \hline
ASBP		 & $1.96$ & $1.93$ & $1.92$ 
\\ \hline
CEMB		 & $1.98$ & $1.94$ & $1.62$ 
\\ \hline
CSBP		 & $1.94$ & $1.89$ & $1.77$ 
\\ \hline
\end{tabular}
    \caption{Convergence rates of ${\cal E}(g_{yy})$ for the 2D
             linearized wave test.}
    \label{tab:lwconv2D}
  \end{center}
\end{table}

\subsection{Gauge wave tests}

For the 1D gauge wave tests, Table~\ref{tab:gwconv1D} shows the convergence
rates of the SBP and embedded algorithms, measured at $t=10$ and $t=50$
crossing times. Second order convergence at $t=10$ is clean in both  cases. 
At $t=50$, the convergence rates of the embedded algorithms show
slight deterioration, for the reasons explained below.

\begin{figure}[hbtp]
  \centering
  \psfrag{xlabel}{t}
  \psfrag{ylabel}{${\cal E}(g_{xx})$}
  \includegraphics*[width=7cm]{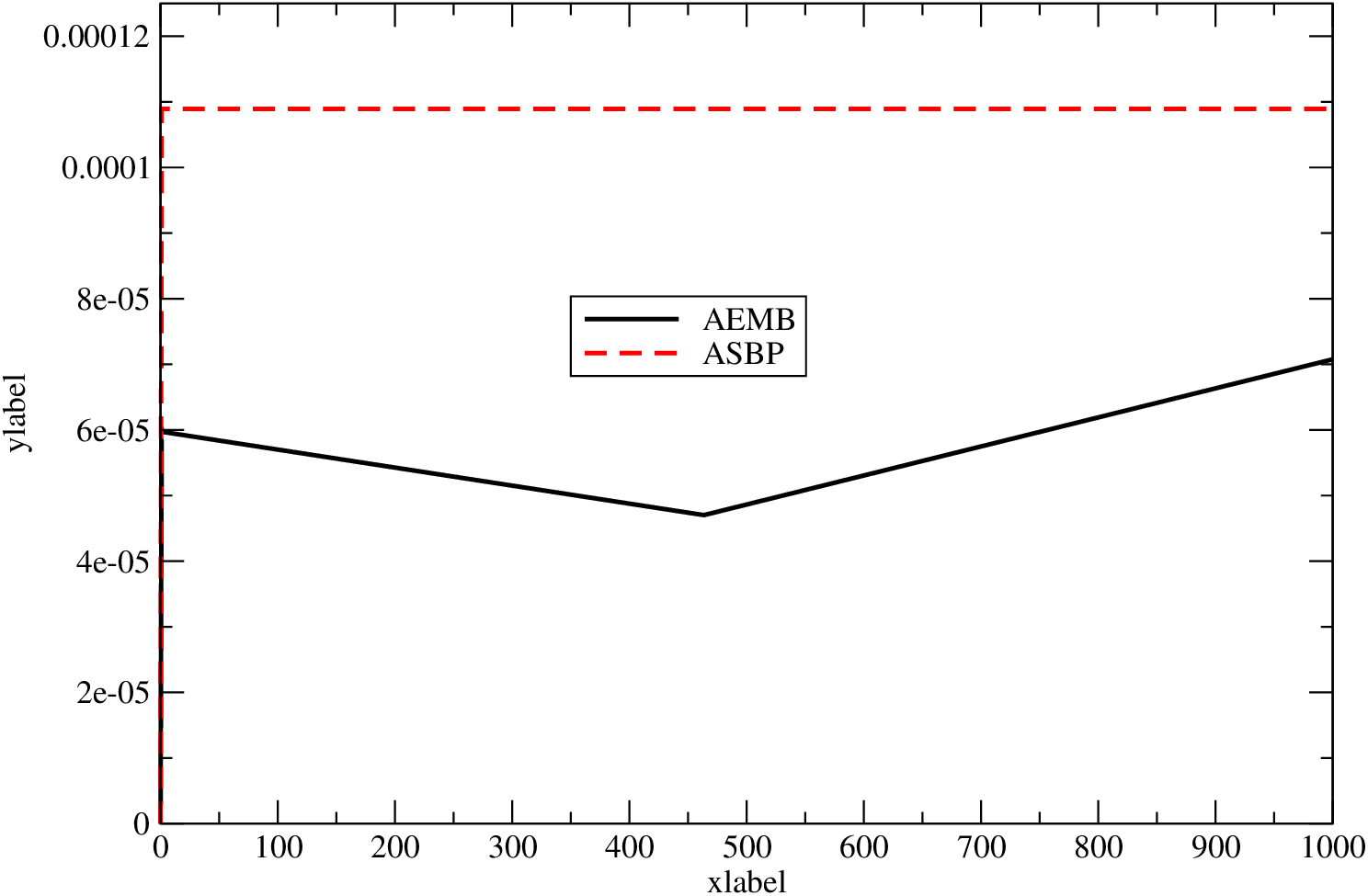}
  \psfrag{xlabel}{x}
  \psfrag{ylabel}{error in $g_{xx}$}
  \includegraphics*[width=7cm]{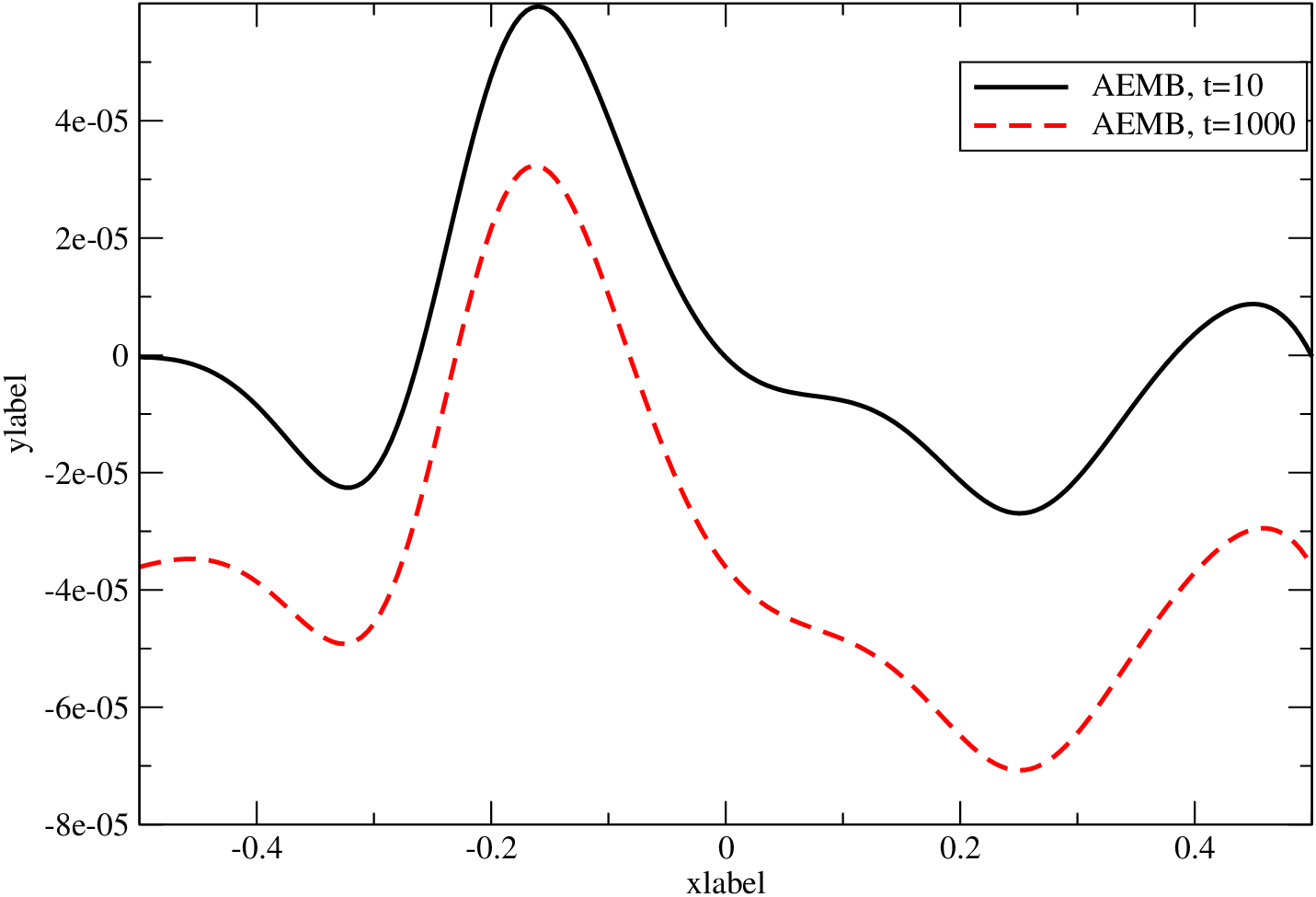}
  \caption{The performance of the analytic
  SBP and embedded algorithms for the 1D gauge wave test. 
  On the left, the error norm ${\cal E}(g_{xx})$ is plotted vs $t$. 
  On the right, snapshots of the error in $g_{xx}$ for the AEMB
  algorithm are shown at $t=10$ and $t=1000$. 
  This error is long wavelength, consisting of two peaks.
  The positive peak dominates in the beginning, but an overall drop in
  the profile causes the error to pass through a minimum and then to grow
  again as the negative peak dominates. This behavior is due to the
  long wavelength instability in the gauge wave spacetime.}
  \label{fig:Agwlt}
\end{figure}

All the algorithms remain stable for 1000 crossing times for the 1D runs on the
$\rho=4$ grid, without use of constraint adjustments or artificial dissipation
except for a small amount of dissipation, (\ref{eq:dissT}) with $\epsilon_T =
0.01$, for  the constrained CSBP algorithm. Figure~\ref{fig:Agwlt} shows the long
term performance of the ASBP and AEMB algorithms. Both maintain excellent
accuracy for 1000 crossing times. For the ASBP algorithm there is negligible
error growth. For the embedded AEMB algorithm, there is long wavelength error
corresponding to the harmonic instability~\cite{bab}
\begin{equation}
  ds_\lambda^2=e^{\lambda t}(1-H)(- dt^2 +dx^2)+dy^2+dz^2 .
\label{eq:lgw}
\end{equation}
of the gauge wave spacetime (\ref{eq:gw}). For any value of $\lambda$,
(\ref{eq:lgw}) is a flat metric which obeys the harmonic constraints. As
depicted in Fig.~\ref{fig:Agwlt}, the resulting profile of the AEMB error
contains two peaks. The positive peak dominates in the beginning, but there is an
overall drop in the waveform, due to the instability, which causes the error to
pass through a minimum and then grow again as the negative peak dominates. The
SBP algorithm is designed to suppress this instability by means of discrete
conservation laws for the principle part of the evolution equations. The embedded
algorithm excites the instability although at a fairly innocuous level.

\begin{figure}[hbtp]
  \centering
  \psfrag{xlabel}{t}
  \psfrag{ylabel}{${\cal E}(g_{xx})$}
  \includegraphics*[width=7cm]{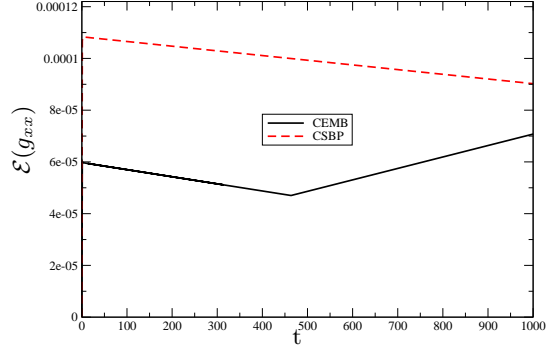}
  \caption{Plot of the error norm ${\cal E}(g_{xx})$ {\it vs} $t$ for the 1D gauge
   wave simulations with the constrained SBP and embedded algorithms,
   obtained with the $\rho=4$ grid.}
  \label{fig:Cgwlt}
\end{figure}

\begin{figure}[hbtp]
  \centering
  \psfrag{xlabel}{t}
  \psfrag{ylabel}{${\cal E}(g_{xx})$}
  \includegraphics*[width=7cm]{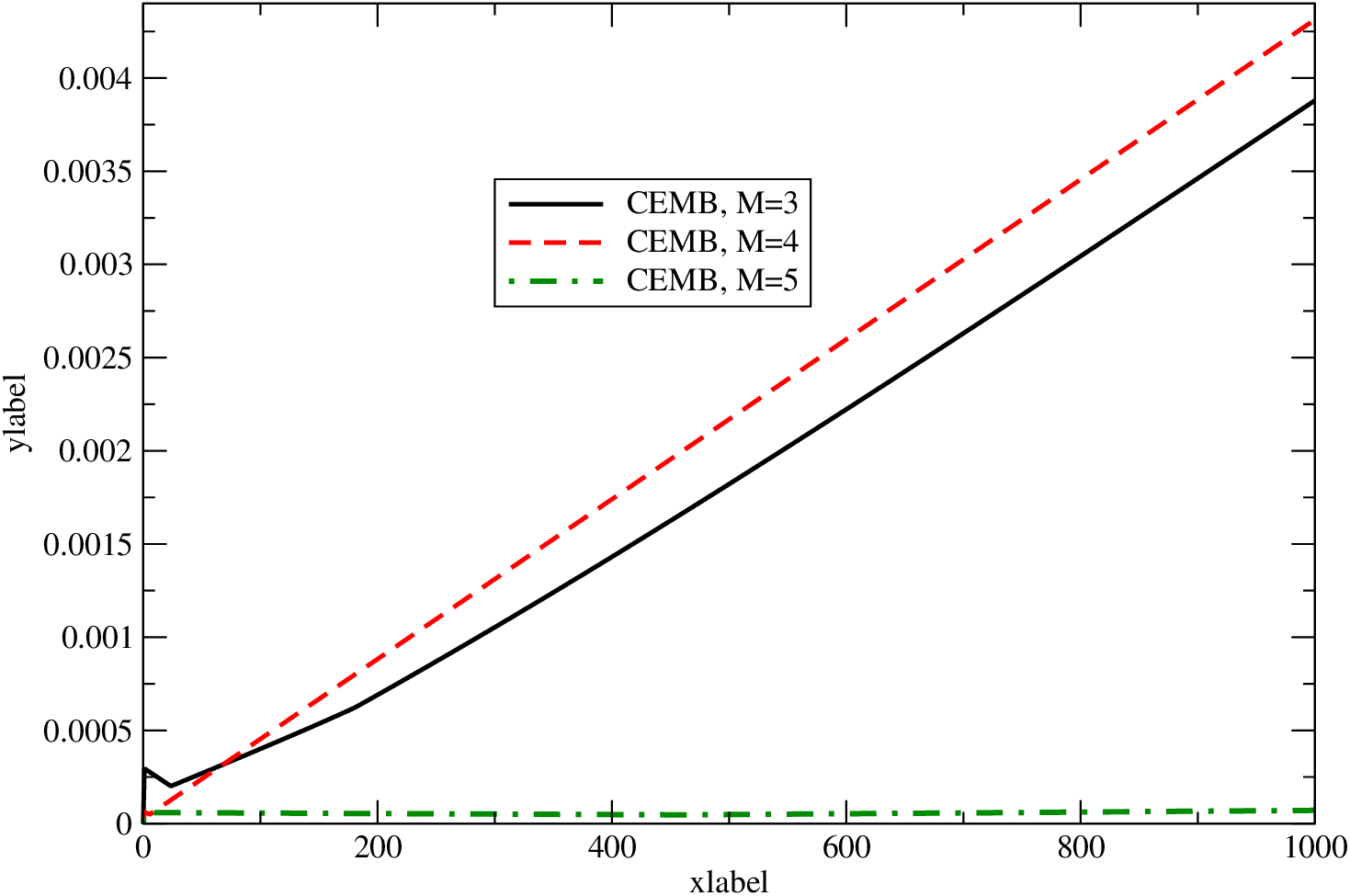}
  \psfrag{xlabel}{x}
  \psfrag{ylabel}{error in $g_{xx}$}
  \includegraphics*[width=7cm]{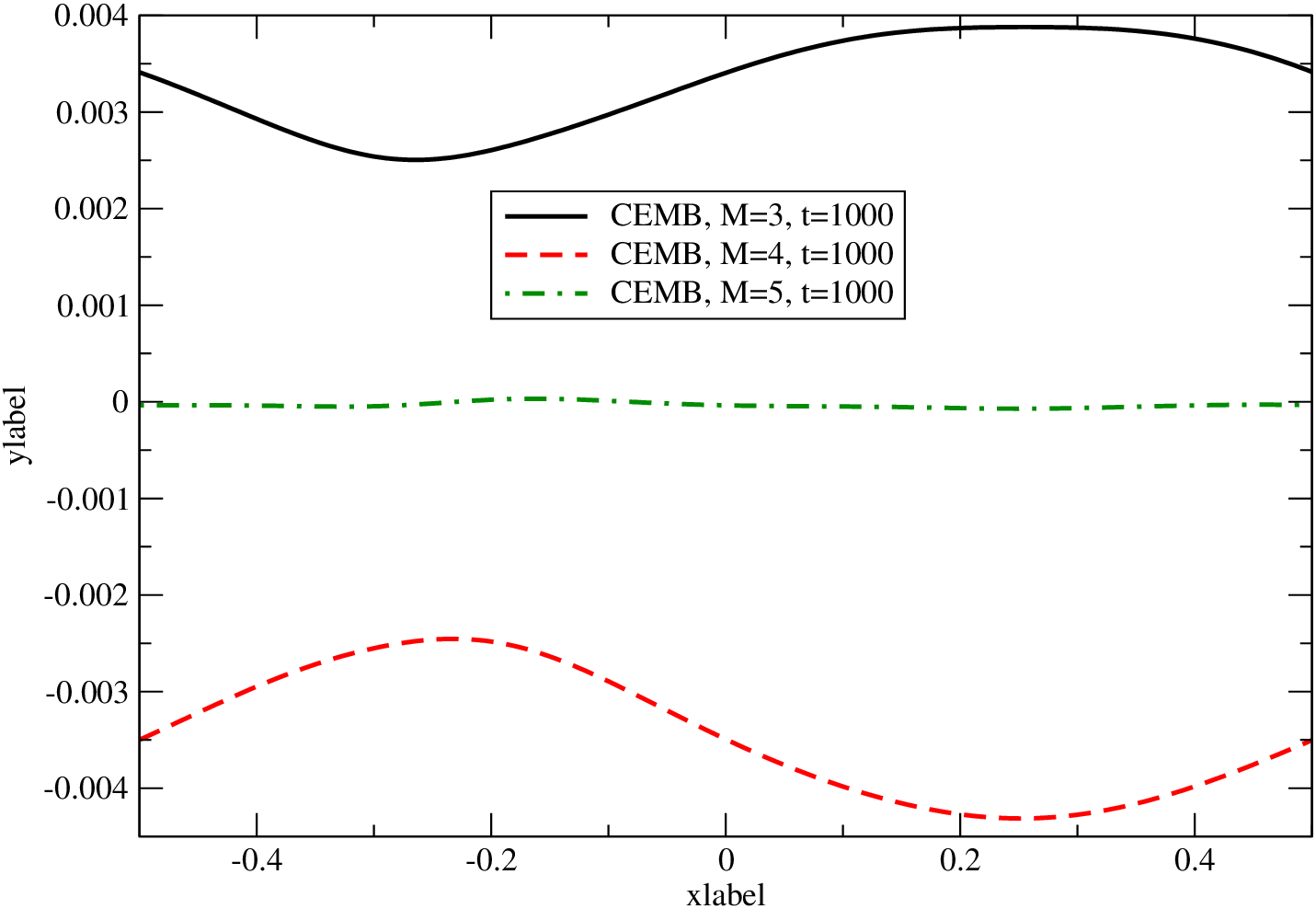}
  \caption{On the left, the error norm ${\cal E}(g_{xx})$ is plotted vs $t$ for
           the 1D gauge wave test using the CEMB algorithm with Lagrange
           polynomials based upon $M=3$, $4$ and $5$ points. Only for $M=5$
           is there no long term error growth. On the right, the snapshots
           of the error in $g_{xx}$ at $t=1000$ show the long wavelength mode
           which is excited in the $M=3$ and $4$ cases.
           }
  \label{fig:CEMBgwlt}
\end{figure}

Similar 1D results hold for the constrained algorithms CSBP and CEMB, as shown in
Fig.~\ref{fig:Cgwlt}. The use of a 5 point $(M=5)$ Lagrange polynomial is
essential for the good performance of the constrained algorithms.
Figure~\ref{fig:CEMBgwlt} shows the rapid error growth which would result from
the use of 3 or 4 points, again arising from excitation of the long wavelength
instability (\ref{eq:lgw}). 

\begin{table}[htbp]
  \begin{center}
    \begin{tabular}[c]{|c|c|c|}
\hline
  ALGORITHM &
  t=10 &
  t=50 
   
    \\ \hline \hline
AEMB		 & $1.97$ & $1.87$ 
\\ \hline
ASBP		 & $2.00$ & $2.00$ 
\\ \hline
CEMB		 & $1.97$ & $1.87$ 
\\ \hline
CSBP		 & $1.98$ & $1.95$ 
\\ \hline
\end{tabular}
    \caption{Convergence rates of ${\cal E}(g_{xx})$ for the 1D
             gauge wave test.
          }
    \label{tab:gwconv1D}
  \end{center}
\end{table}

\begin{figure}[hbtp]
  \centering
  \psfrag{xlabel}{t}
  \psfrag{ylabel}{${\cal E}(g_{xx})$}
  \includegraphics*[width=7cm]{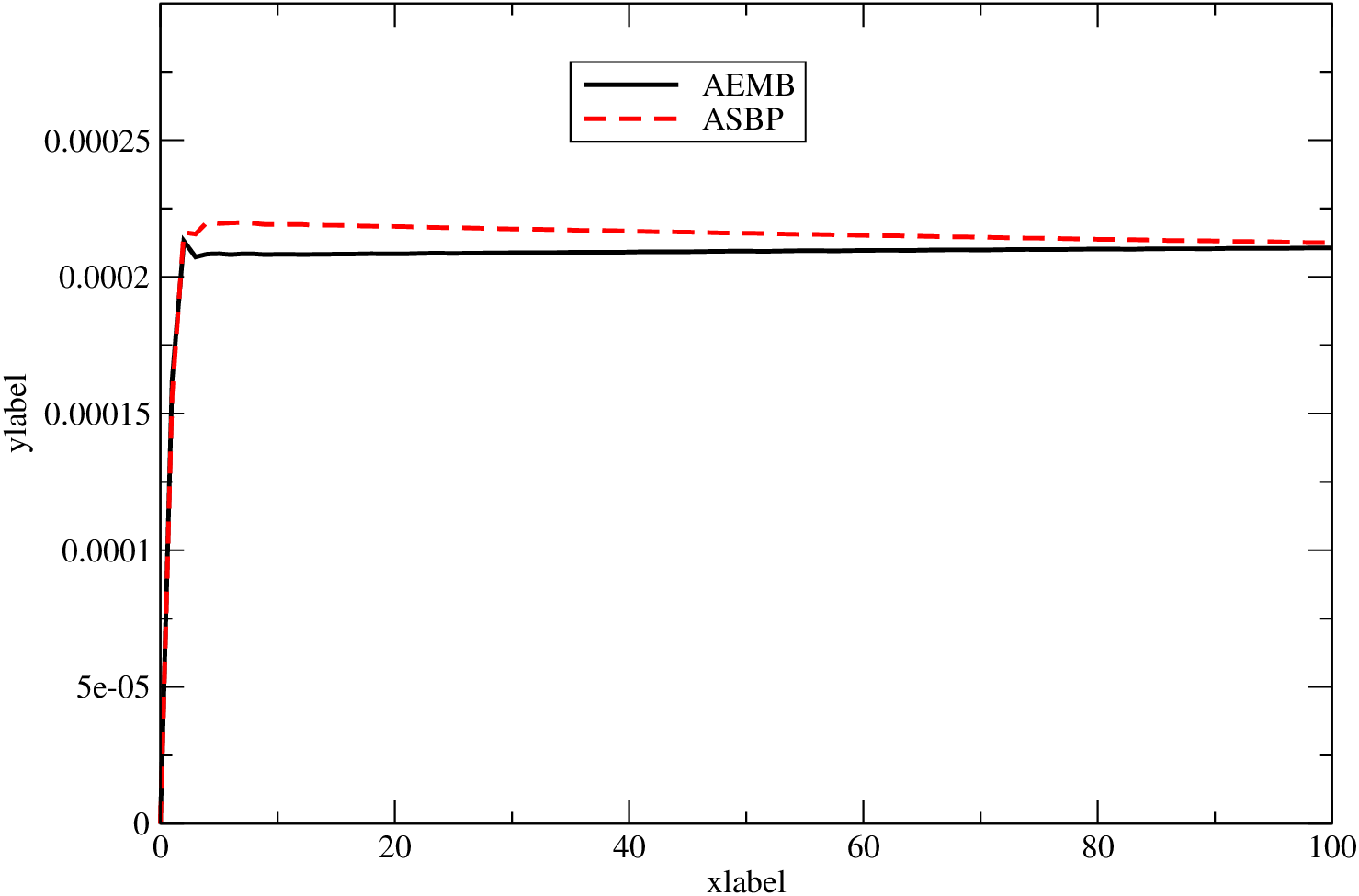}
  \includegraphics*[width=7cm]{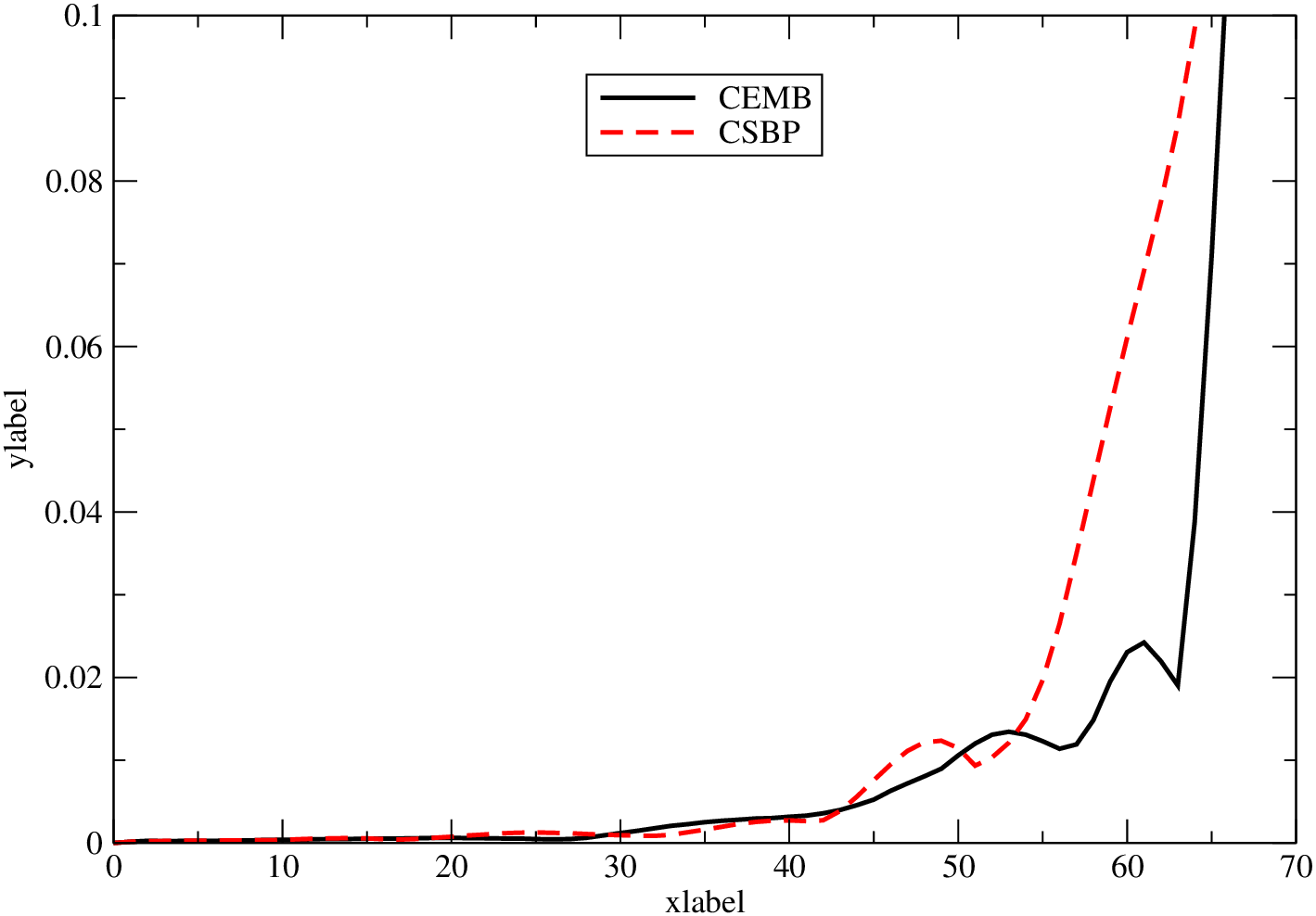}
  \caption{Plots of ${\cal E}(g_{xx})$ {\it vs} $t$ for the 2D gauge wave tests.
   The left graphs compare the error for  the analytic SBP and embedded
   algorithms. The right graphs, which compare the error in the constrained SBP
   and embedded algorithms, exhibit the excitation of the long wavelength
   instability.}
  \label{fig:2Dgwlt}
\end{figure}

The convergence rates for the 2D gauge wave tests shown in
table~\ref{tab:gwconv2D} indicate clean second order convergence up to $t=10$.
The graphs of the error in Fig.~\ref{fig:2Dgwlt} show that the analytic SBP and
embedded algorithms maintain excellent accuracy up to $t=100$. However, the
constrained algorithms excite the long wavelength instability (\ref{eq:lgw}) at
$t\approx 55$ for CSBP and $t\approx 60$ for CEMB. Neither numerical
dissipation nor constraint adjustment lead to significant improvement.
Higher order accuracy of the boundary condition also does not seem to help.

\begin{table}[htbp]
  \begin{center}
    \begin{tabular}[c]{|c|c|c|c|}
\hline
  ALGORITHM &
  t=1 &
  t=5 &
  t=10 
   
    \\ \hline \hline
AEMB		 & $2.02$ & $2.03$ & $2.03$ 
\\ \hline
ASBP		 & $2.02$ & $2.03$ & $2.02$ 
\\ \hline
CEMB		 & $2.02$ & $2.01$ & $2.02$ 
\\ \hline
CSBP		 & $2.02$ & $2.01$ & $2.03$ 
\\ \hline
\end{tabular}
    \caption{Convergence rates of ${\cal E}(g_{xx})$ for the 2D gauge
            wave test.
          }
    \label{tab:gwconv2D}
  \end{center}
\end{table}

\subsection{Shifted gauge wave tests}

For the 1D shifted gauge wave tests, Table~\ref{tab:sgwconv} shows the
convergence rates of the SBP and embedded algorithms measured at $t=10$ and
$t=50$ crossing times. Second order convergence is fairly clean at $t=10$, with
some drifting of the 
convergence rates evident at $t=50$. 

\begin{table}[htbp]
  \begin{center}
    \begin{tabular}[c]{|c|c|c|}
\hline
  ALGORITHM &
  t=10 &
  t=50 
    \\ \hline \hline
AEMB		 & $1.97$ & $1.89$ 
\\ \hline
ASBP		 & $2.08$ & $2.34$ 
\\ \hline
CEMB		 & $1.90$ & $1.97$
\\ \hline
CSBP		 & $2.08$ & $2.25$ 
\\ \hline
\end{tabular}
    \caption{Convergence rates of ${\cal E}(g_{xx})$ for the 1D shifted gauge
      wave test.
            }
    \label{tab:sgwconv}
  \end{center}
\end{table}

\begin{figure}[hbtp]
  \centering
  \psfrag{xlabel}{t}
  \psfrag{ylabel}{${\cal E}(g_{xx})$}
  \includegraphics*[width=7cm]{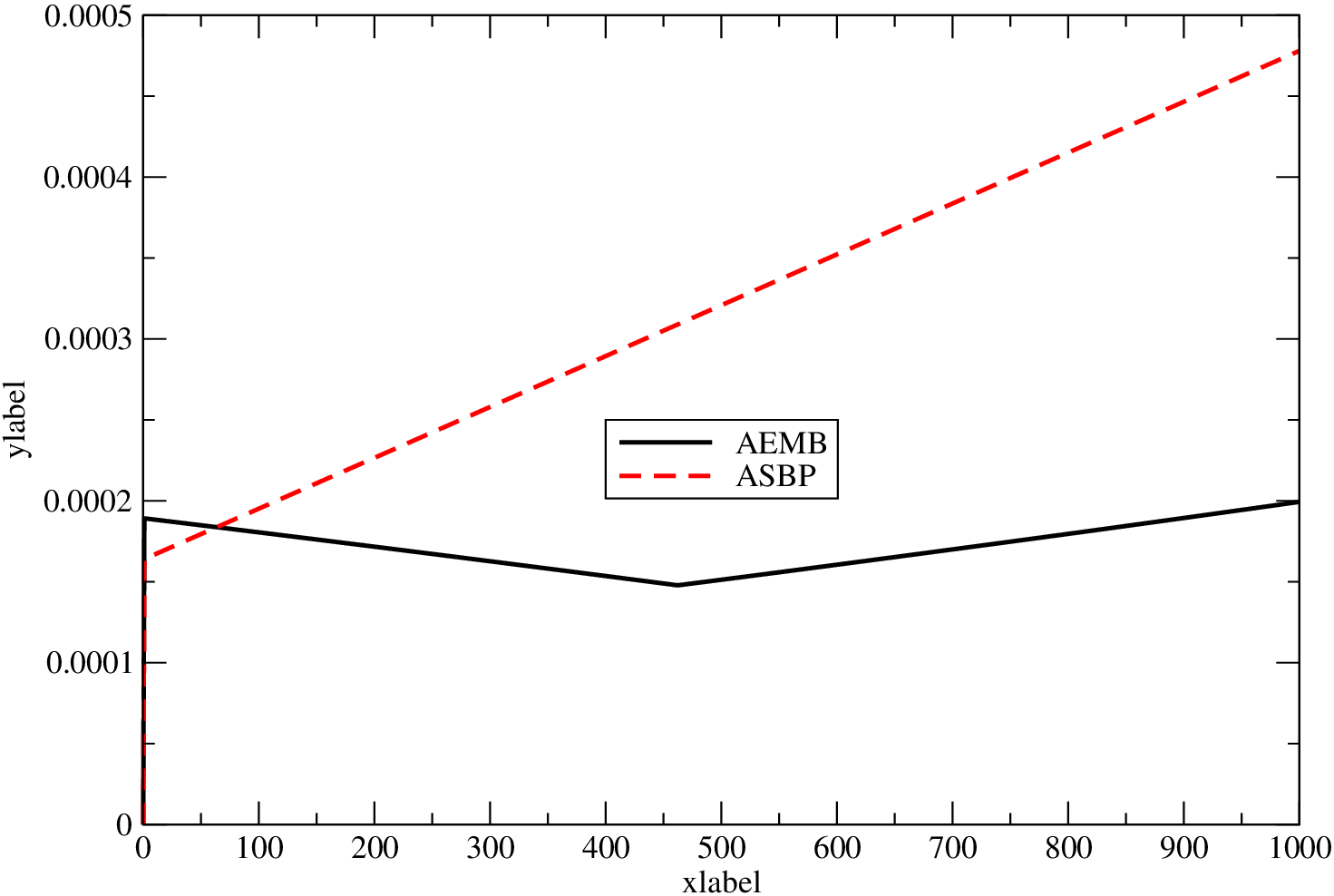}
  \psfrag{xlabel}{x}
  \psfrag{ylabel}{error in $g_{xx}$}
  \includegraphics*[width=7cm]{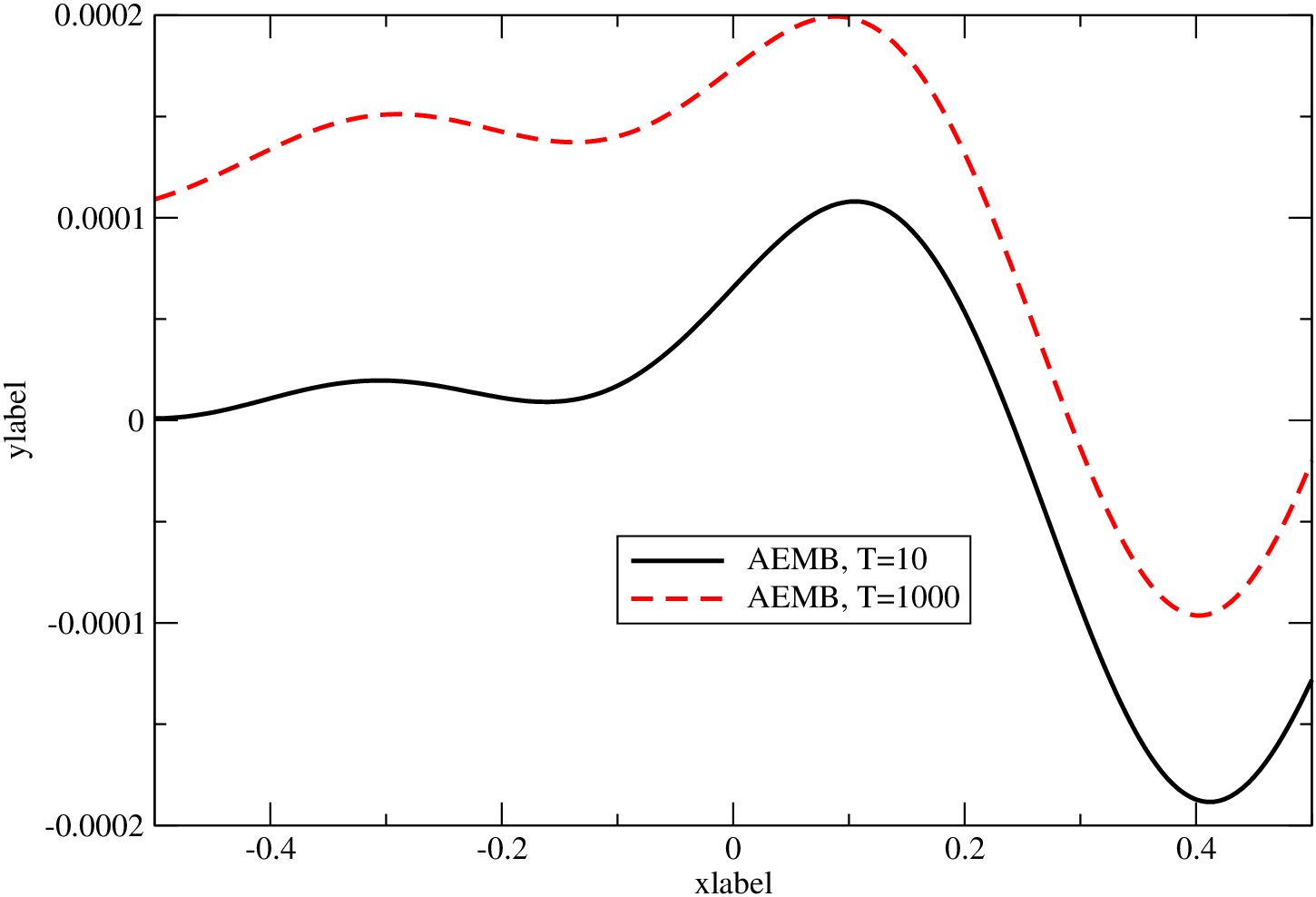}
  \caption{Performance of the analytic SBP and embedded algorithms
   for the 1D shifted gauge wave. On the left, the error norm ${\cal E}(g_{xx})$ 
   is plotted {\it vs} $t$. On the right, snapshots of the error for the AEMB
   algorithm with the $\rho=4$ grid are shown at $t=10$ and $t=1000$. The error
   is long wavelength,
   consisting of two peaks. The overall rise in the profile is due to low level
   excitation of the long wavelength instability (\ref{eq:instab}).}
  \label{fig:Asgwlt}
\end{figure}

A long wavelength instability also exists in the shifted gauge wave 
spacetime
(\ref{eq:sgw})~\cite{babev},
\begin{equation}
  \label{eq:lsgaugewave4metric}
  ds_\lambda^2=- dt^2 +dx^2+dy^2+dz^2 +
        \bigg(H-1+e^{\lambda \hat t}\bigg )
            k_\alpha k_\beta dx^\alpha dx^\beta ,
\label{eq:instab}
\end{equation}
where
\begin{equation}
\hat t= t - \frac {Ad}{4\pi}\cos \left( \frac{2 \pi (x - t)}{d} \right) .
\end{equation}
Although this metric does not solve Einstein's
equations, for any value of $\lambda$ it satisfies the standard harmonic form
(\ref{eq:e}) of the reduced Einstein equations, i.e. the equations
governing numerical evolution without constraint adjustment.
This instability is the major source of error.
Figure~\ref{fig:Asgwlt} exhibits the long term performance for runs with the
analytic SBP and embedded algorithms. They maintain
excellent accuracy for 1000 crossing times, although the snapshots show that
the embedded algorithm has produced a low level excitation of the
instability.

\begin{figure}[hbtp]
  \centering
  \psfrag{xlabel}{t}
  \psfrag{ylabel}{${\cal E}(g_{xx})$}
  \includegraphics*[width=7cm]{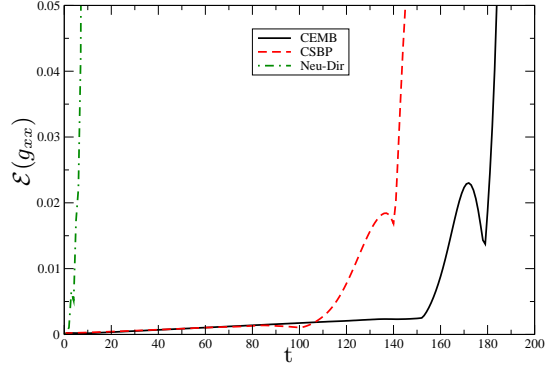}
  \caption{Plot of ${\cal E}(g_{xx})$ vs $t$ for the 1D gauge
   wave with shift simulations for the constrained CEMB and CSBP algorithms, 
   with numerical dissipation $\epsilon_T=0.001$ and constraint adjustment
   with $b=1$, and for the Neumann-Dirichlet constrained preserving algorithm.
   The comparison shows that both Sommerfeld constrained algorithms 
   clearly outperform the Neumann-Dirichlet constrained algorithm.}
  \label{fig:Csgwlt}
\end{figure}

\begin{figure}[hbtp]
  \centering
  \psfrag{xlabel}{t}
  \psfrag{ylabel}{${\cal E}(g_{xx})$}
  \includegraphics*[width=7cm]{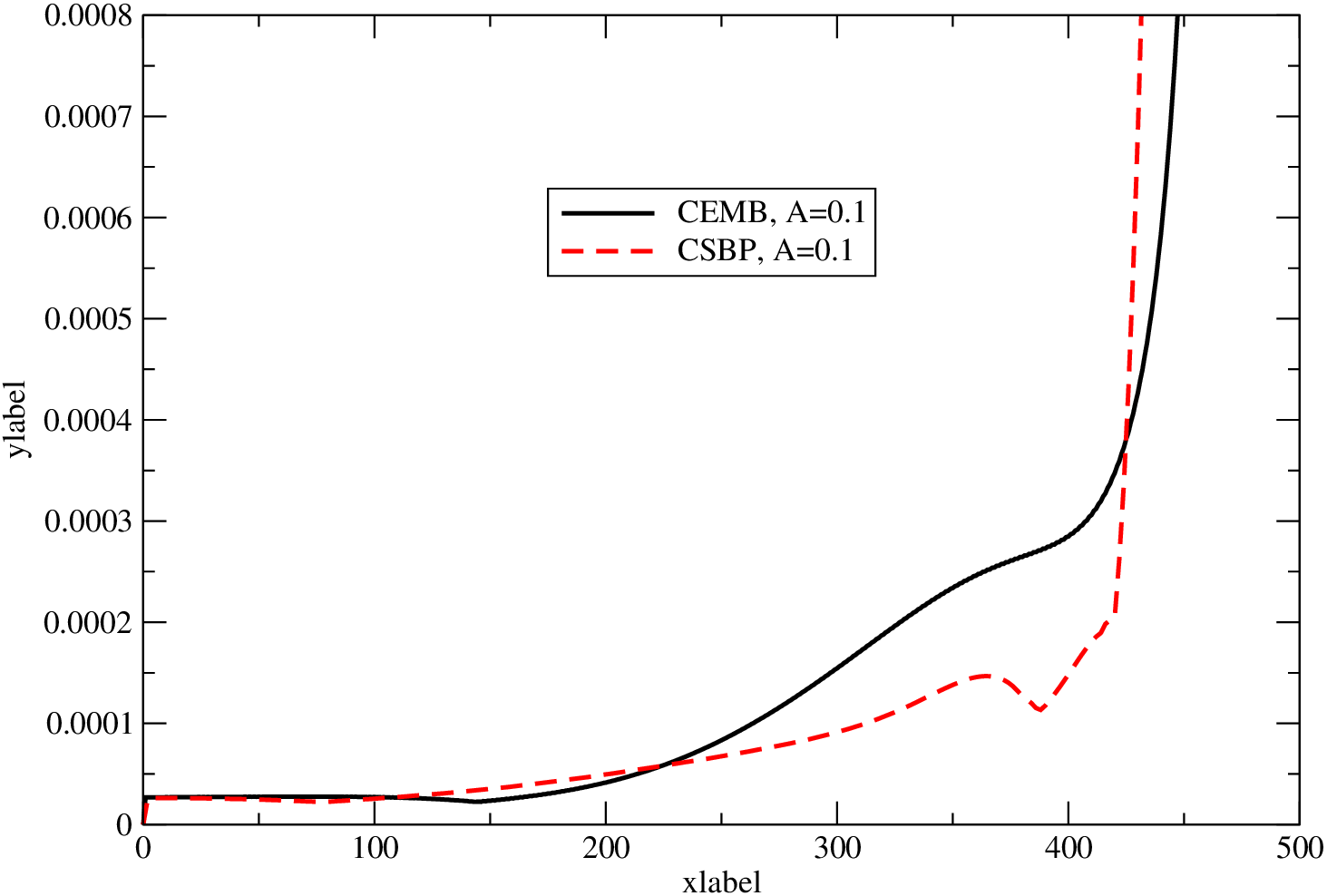}
  \includegraphics*[width=7cm]{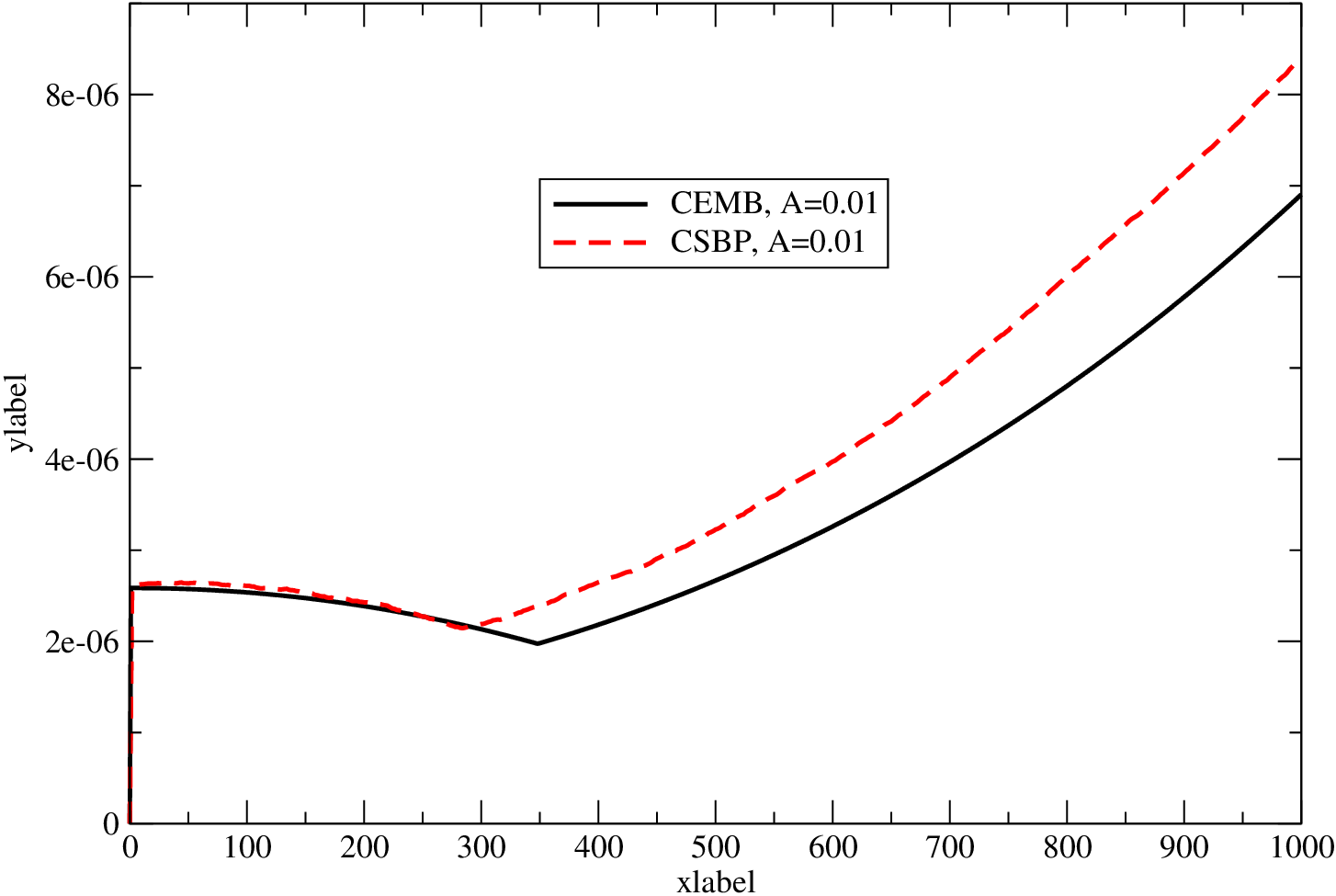}
  \caption{Plots of ${\cal E}(g_{xx})$ vs $t$ for lower amplitude simulations
  of the 1D gauge wave with shift simulations for the
  constrained versions of the SBP and embedded algorithms. The nonlinear nature
  of the stability is evident.
   }
  \label{fig:lincomp}
\end{figure}

The larger boundary errors of the constrained algorithms leads to stronger
excitation of the long wavelength instability. Without constraint adjustment,
the runs on the $\rho=4$ grid crash at
about $t\approx 88$  for the CSBP algorithm and $t\approx 103$ for the CEMB
algorithm.   
Numerical dissipation was necessary in the CEMB run but had insignificant
effect on controlling the long wavelength instability in either
of the constrained algorithms.
However, constraint adjustment was moderately effective. Of the various
adjustments of the form (\ref{eq:constadj}) which were considered in~\cite{babev},
the longest constrained runs were obtained with the choice
\begin{equation}
      A^{\mu\nu}= \frac{b {\cal C}^\alpha \nabla_\alpha t}
      {e_{\rho\sigma}{\cal C}^\rho {\cal C}^\sigma}
         {\cal C}^{\mu} {\cal C}^{\nu}, \quad  b>0 .
\label{eq:mmmmbeladj}
\end{equation}
Here 
\begin{equation}
        e_{\rho\sigma}=g_{\rho\sigma}- 
          \frac{2}{g^{tt}}(\nabla_\rho t)\nabla_\sigma t
\end{equation}
is the natural metric of signature $(++++)$ associated with the Cauchy
slicing. This adjustment with $b=1$ extended the CSBP run to $t\approx 155$
and the CEMB run to  $t\approx 183$ crossing times, as indicated in 
Figure~\ref{fig:Csgwlt}. The response to constraint adjustment is evidence of
the constraint violating origin of the analytic instability. The instability
is also of nonlinear origin, which can be seen from the comparison with the
runs of lower amplitude shown in Fig.~\ref{fig:lincomp}. With lower amplitude,
the discrete conservation laws of the SBP algorithm begin to control the
instability and give performance comparable to the CEMB algorithm. As was
previously found for shifted gauge wave tests with periodic boundary
conditions~\cite{babev}, constraint damping~\cite{constrdamp} introduces 
oscillations with unacceptably large error and does not appreciably suppress
the instability. Nevertheless, the performance of the constrained
Sommerfeld algorithms is vastly superior to the Neumann-Dirichlet
constrained algorithm considered in~\cite{harl}, as can be seen in
figure~\ref{fig:Csgwlt}

\begin{figure}[hbtp]
  \centering
  \psfrag{xlabel}{t}
  \psfrag{ylabel}{${\cal E}(g_{xx})$}
  \includegraphics*[width=7cm]{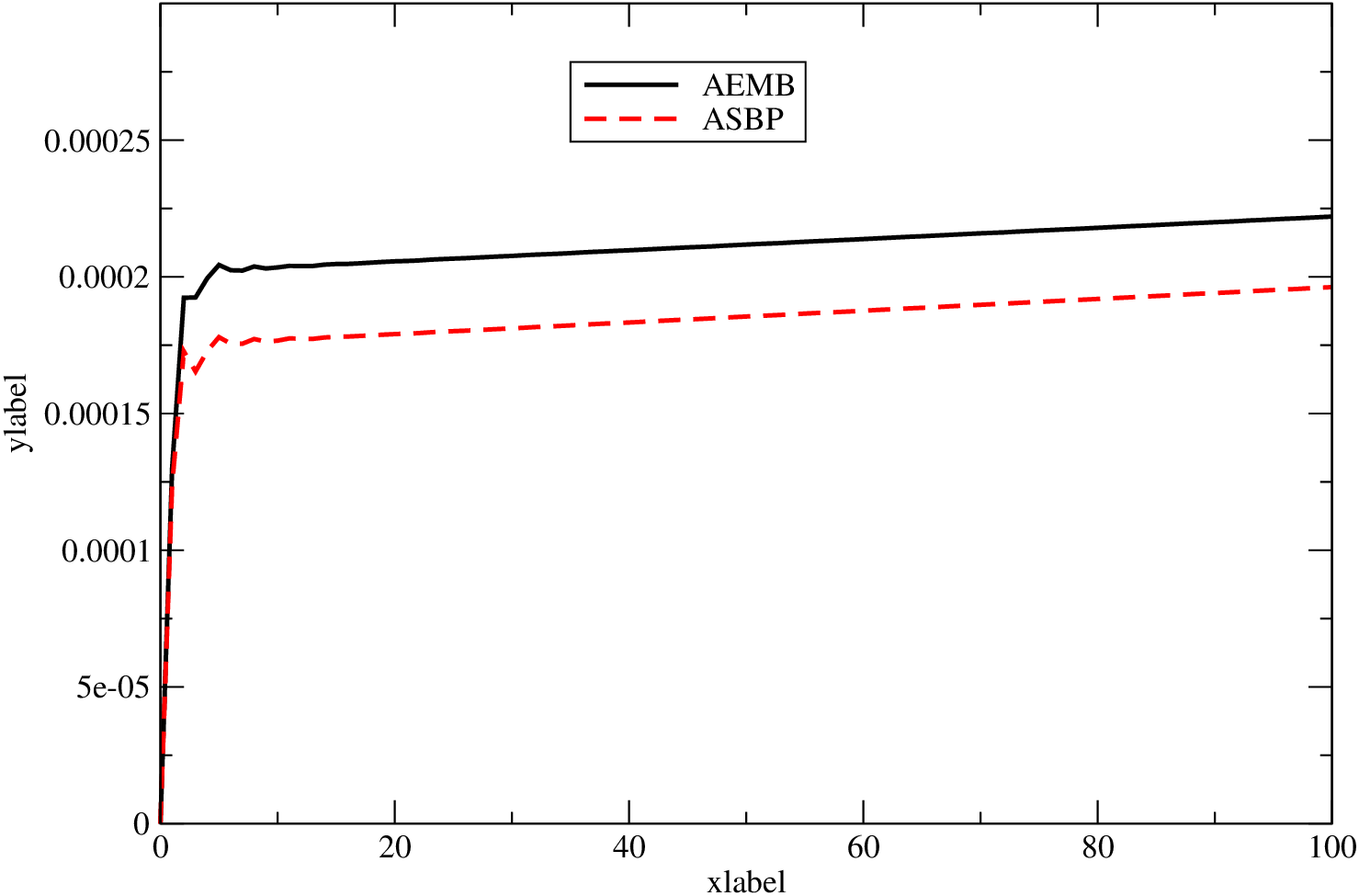}
  \includegraphics*[width=7cm]{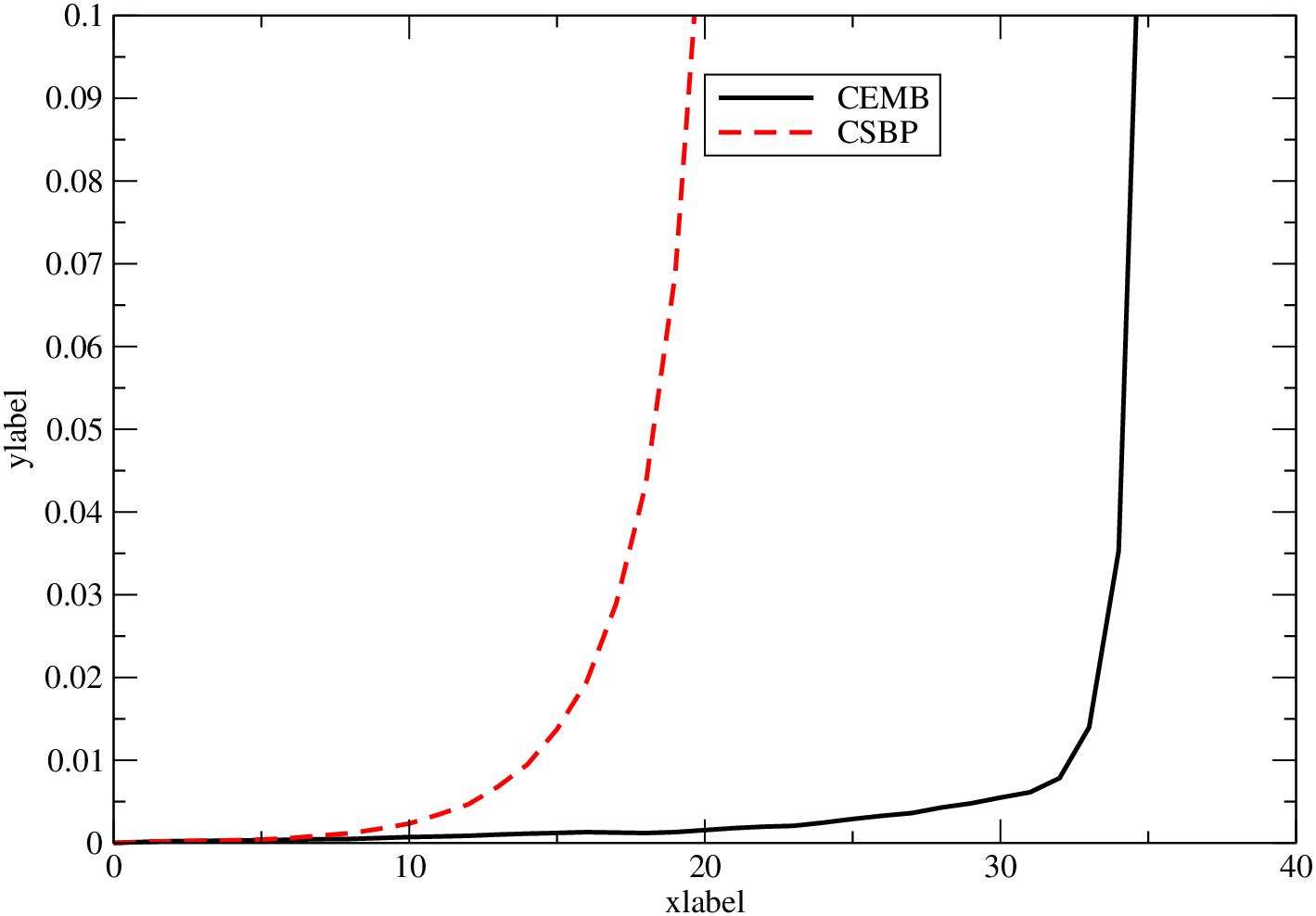}
  \caption{Plots of ${\cal E}(g_{xx})$ vs $t$ for the 2D gauge wave 
   with shift. The left graphs compare the 
   analytic SBP and embedded algorithms and the right graphs compare
   the constrained algorithms.}
  \label{fig:2Dsgwlt}
\end{figure}

The convergence rates for the 2D shifted gauge wave tests shown in
table~\ref{tab:sgwconv2} indicate clean second order convergence up  to $t=10$.
Figure~\ref{fig:2Dsgwlt} exhibits the long term performance for runs with the
analytic and  constrained SBP and embedded algorithms.
For the analytic algorithms (left plot),  both the ASBP and the
AEMB algorithms maintain excellent accuracy  for 1000 crossing times. This is
one of the few cases where the error in  the embedded algorithm is larger than
the  error in the SBP algorithm. This results from the  2D interpolation error
introduced by the embedded algorithm.  For the constrained algorithms (right
plot), the long  wavelength instability is excited earlier by the CSBP
algorithm, which crashes at $t\approx22$, while the CEMB  algorithm runs up to
$t\approx35$.  Efforts to control the instability by constraint adjustment and
numerical dissipation had insignificant effect in prolonging the runs. 

\begin{table}[htbp]
  \begin{center}
    \begin{tabular}[c]{|c|c|c|c|}
\hline
  ALGORITHM &
  t=1 &
  t=5 &
  t=10 
    \\ \hline \hline
AEMB		 & $2.00$ & $2.01$ & $2.03$ 
\\ \hline
ASBP		 & $2.00$ & $2.02$ & $2.05$
\\ \hline
CEMB		 & $2.00$ & $2.00$ & $1.97$
\\ \hline
CSBP		 & $2.00$ & $2.00$ & $2.04$
\\ \hline
\end{tabular}
    \caption{Convergence rate of ${\cal E}(g_{xx})$ for the 2D shifted gauge
      wave test.
            }
    \label{tab:sgwconv2}
  \end{center}
\end{table}    

\vfill\eject

\bigskip

\section{Conclusion}

The preceding tests involved linearized wave, gauge wave and shifted gauge wave
metrics for which the exact solutions provide the correct boundary data. The
results provide some definitive conclusions. First, the results show that
constraint-preserving Sommerfeld boundary conditions give good long term
accuracy, which is vastly superior to previous shifted gauge wave test
results for constraint-preserving  Dirichlet-Neumann conditions. 

Of foremost importance, the analytic versions of both the SBP and embedded
algorithms clearly outperformed the constrained versions. The knowledge of the
exact Sommerfeld boundary data avoids the need for computing
constraint-preserving boundary conditions. One might have expected numerical
noise to generate constraint violating error which the analytic algorithms
could not properly handle. While this undoubtedly occurs, the analytic
algorithms nevertheless outperform the constrained algorithms because the prime
source of error is due to the long wavelength instabilities inherent in the
nonlinear test problems. The additional error introduced by computing
constrained Sommerfeld data leads to earlier excitation of these long
wavelength instabilities. 

Also of importance, our tests results show no clear advantage of either the
SBP or embedded treatments of the boundary condition. In the more complicated
case of a curved boundary, both approaches are being developed
by their advocates in the computational mathematics community.

In a realistic problem, such as the binary black hole problem, a global scheme
is necessary to provide physically correct outer boundary data, either by using
hyperboloidal time slices to extend the Cauchy evolution to infinity
(see~\cite{Hhprbloid,joerg} for reviews) or by matching to an exterior
characteristic or perturbative solution (see~\cite{winrev} for a review).
Harmonic evolution offers important advantages for Cauchy-characteristic
matching (CCM) which have led to successful matching in the linearized
regime~\cite{harl}. Primarily, the harmonic constraints can be enforced by
propagating the Cauchy coordinates on the characteristic grid. This provides
the proper Jacobian for injecting the Cauchy boundary data from the
characteristic solution. The results of this paper should supply helpful
guidance for extending the harmonic version of CCM to the nonlinear problem.  

\centerline{\bf Acknowledgments} 

We have benefited from the use of the Cactus Computational Toolkit
(http://www.cactuscode.org). We thank Ian Hawke for his help in utilizing the
Cactus boundary routines. Computer time was provided by the Pittsburgh
Supercomputing Center through a TeraGrid Wide Roaming Access 
Computational Resources Award. This work was supported by the National Science
Foundation under grant PH-0553597 to the University of Pittsburgh.

\end{document}